\documentstyle[prd,eqsecnum,aps,epsf]{revtex}

\newcommand{\singlefig}[2]{
\begin{center}
\begin{minipage}{#1}
\epsfxsize=#1
\epsffile{#2}
\end{minipage}
\end{center}}
\newcommand{\segmentfig}[3]{
\begin{minipage}{#1}
\epsfxsize=#1
\epsffile{#2}
\begin{center}
{\small \mbox{#3}}
\end{center}
\end{minipage}}
\newcommand{\lsim}{\mbox{\raisebox{-1.ex}{$\stackrel
     {\textstyle<}{\textstyle \sim}$}}}
\begin{document}
\draft
\title{Gravitating monopole and its black hole solution in Brans-Dicke 
Theory}
\author{Takashi Tamaki\thanks{electronic
mail:tamaki@gravity.phys.waseda.ac.jp}
and Kei-ichi Maeda\thanks{electronic
mail:maeda@gravity.phys.waseda.ac.jp}
}
\address{Department of Physics, Waseda University,
Shinjuku, Tokyo 169, Japan}
\author{Takashi Torii\thanks{electronic
mail:torii@th.phys.titech.ac.jp}
}
\address{Department of Physics, Tokyo Institute of
Technology, Meguro, Tokyo 152, Japan}
\date{\today}
\maketitle
\begin{abstract}
We find a self-gravitating monopole and its black hole solution 
in Brans-Dicke (BD) theory. 
We mainly discuss the properties of these solutions 
in the Einstein frame and compare the solutions 
with those in general relativity (GR) on the following points. : 
From the field distributions of the generic type of self-gravitating monopole solutions, 
we can find that the YM potential and 
the Higgs field hardly depend on the BD parameter for most of the solution. 
There is the upper limit of the vacuum expectation value of the 
Higgs field to exist a solution as in GR. 
Since the BD scalar field has the effect of lessening an effective gauge charge, 
the upper limit in BD theory (in the $\omega =0$ case) becomes about $30\%$ 
larger than in GR. 
In some parameter ranges, there are two nontrivial solutions with the
same mass, one of which can be regarded as the excited state of the other.
This is confirmed by the analysis by catastrophe theory,
which stated that the excited solution is unstable.
We also find that the BD scalar field varies larger for solutions of smaller horizon 
radius, which can be  understood from the differences of non-trivial structure outside the horizon. 
A scalar mass and thermodynamical properties of new solutions are 
also examined.
Our analysis may give an insight into solutions in other theories of gravity; particularly, a theory 
with a  dilaton field may show similar effects because of its coupling to a gauge field. 
\end{abstract}

%%%%%%%%%%%%%%%%%%%%%%%%%%%%%%%%%%%%%%%%%%%%%%%%%%
%
%                                                 %
\section{Introduction}                            %
%                                                 %
%%%%%%%%%%%%%%%%%%%%%%%%%%%%%%%%%%%%%%%%%%%%%%%%%%
After the discovery of a particlelike solution in 
the Einstein-Yang-Mills (EYM) system by Bartnik and Mckinnon \cite{BM}, 
a variety of particlelike solutions and black hole solutions 
with non-Abelian fields were
found\cite{Tor2,rf:colored,Torii,DHS,LM,Tor,rf:Greene,Lee,Ortiz,BFM,Aichelburg,Tachi}. 
Some of them are the solutions in which the self-gravity is taken 
into account for the solution in  flat space-time. The others are new types 
of solutions which can exist in the presence of gravity.
They show interesting properties, which can not 
be seen in the flat spacetime solutions or well known Kerr-Newman black hole, such as stability, thermodynamical 
properties, mass inflation phenomena and so on\cite{Donets}. 
In the context of a counterexample for the black hole no-hair conjecture, 
a monopole black hole\cite{Lee,Ortiz,BFM,Aichelburg,Tachi}
may be the most important among them, since both the Reissner-Nortstr\"{o}m (RN) 
black hole and the monopole black hole with the same mass and the gauge charge are stable within a certain 
parameter region and it is constructed in the EYM-Higgs (EYMH) system which is not a phenomenological model 
such as the Skyrme and the Proca model. 
The monopole black hole solution describes a space-time structure with 
an event horizon in the gravitating  
't Hooft-Polyakov magnetic monopole\cite{'t Hooft}.   
The existence of such a black hole solution was first expected by examining a simple distribution of the energy density, 
where it is almost constant inside a core and decays as $r^{-4}$ outside\cite{Lee}.  Later Breitenlohner {\it et al.} \cite{BFM}, 
Aichelburg \cite{Aichelburg} and we \cite{Tachi} found the ``exact" solutions numerically. 

In the big bang scenario, the 't~Hooft-Polyakov monopole is overproduced during  phase
transitions in the early universe. This is one of the reasons why the inflationary scenario was 
proposed\cite{Sato}.  Although such an 
inflationary scenario was discussed originally in general relativity (GR), 
it turns out that introduction of a scalar field (such as the Brans-Dicke (BD) 
scalar field) can make a great  change in scenarios of the very 
early universe\cite{rf:Linde}. 
Moreover, Vilenkin and Linde proposed that inflation 
occurs inside the 't~Hooft-Polyakov monopole\cite{Vilenkin},
where no fine tuning of the initial value is necessary. Such a monopole 
inflation is also changed qualitatively by introducing a scalar field\cite{Sakai}. 
So properties of the gravitating monopole and its black hole solution may  
be greatly modified in 
scalar-tensor theory compared with in GR.

As for the dilatonic black holes in an effective theory of superstring, it is known that such a scalar field also 
affects many features\cite{GM}.  In BD theory, 
it is also known that the Kerr or Kerr-Newman black hole is 
a unique solution in the vacuum case or in the case with the Maxwell
field,  because of the conformal invariance of the Maxwell field and the
black hole uniqueness theorem\cite{Hawking}. 
Recently, however, we found non-trivial black hole solutions with the non-Abelian field in 
BD theory\cite{Tamaki}. 
They are generalizations of neutral non-Abelian black holes in GR (Proca\cite{rf:Greene} and 
Skyrme black holes\cite{DHS}). 
Then, it is important to study a self-gravitating monopole and its black hole solution 
in BD theory as the charged solution. 
In addition, we have the following reasons to consider the monopole black hole in BD theory. 
(i) For the neutral non-Abelian black holes, we can not find a more stable solution than 
a trivial solution, i.e., Schwarzschild solution even in BD theory.  We expect that 
the monopole black hole can be more stable than a trivial solution, i.e., 
Reissner-Nortstr\"{o}m (RN) solution in the BD-YMH system. 
So it would be more important than the neutral type.  
Particularly, when we discuss the Hawking radiation\cite{Hrad} of the RN black hole in the EYMH system, 
it may evolve into the monopole black hole and eventually into the gravitating monopole, 
which would be a candidate for its remnant. The scalar field 
may cause a serious effect on this evolution. 
(ii) To investigate the stability of the neutral non-Abelian black holes, we can apply  
catastrophe theory even in BD theory\cite{Torii,Tamaki}. 
Using the variables in the Einstein frame, we find that the BD scalar field is 
less important from the perspective of catastrophe theory. However the solution is not evident in 
the charged monopole black hole case. 

This paper is organized as follows. 
We introduce basic $Ans\ddot{a}tze$ and the field equations in the BD-YMH 
system in 
Sec. II.  In Sec. III, we present numerical solutions for the regular self-gravitating monopole. 
In order to see their structure, we show their field configurations and compare them with the 
solution in GR. We also discuss the upper limit 
of the vacuum expectation value of the Higgs field for a solution to exist. 
In Sec. IV, we present the monopole black hole solution. 
In some parameter regions, there exist two types of non-trivial solutions,  
one of which is stable and the other is not. We show that the 
dependence of field configurations on the BD parameter is different in each type. 
This shows the difference of the black hole structures which can not be seen only 
by considering them in GR. We also discuss upper limits 
for the vacuum expectation value of the Higgs field for a static solution to exist in  
black hole cases. The concluding remarks will follow in
Sec. V. Throughout this paper we use  units of
c=$\hbar$=1.  Notations and definitions follow Misner-Thorne-Wheeler \cite{rf:MTW}.

%%%%%%%%%%%%%%%%%%%%%%%%%%%%%%%%%%%%%%%%%%%%%%%%%%
%%%%%%%%%%
%                                                          %
\section{Basic equations}    %
%                                                          %
%%%%%%%%%%%%%%%%%%%%%%%%%%%%%%%%%%%%%%%%%%%%%%%%%%
%%%%%%%%%%

\ The action of BD theory is written as \cite{BD}
%%%%%%%%%%%%%
\begin{eqnarray}
S=\int d^{4}x\sqrt{-g}\left[\frac{1}{2\kappa^{2}}\left(\phi R-
\frac{\omega}
{\phi}\nabla_{\alpha}\phi\nabla^{\alpha}\phi
\right) + L_{m}\right],
\label{B1}
\end{eqnarray}
%%%%%%%%%%%%%
where $\kappa^{2} \equiv 8\pi G$ with $G$ being Newton's
gravitational constant. $\omega$ is  the BD parameter, and
$L_{m}$ is  Lagrangian of the matter fields. The
dimensionless BD scalar field $\phi$ is normalized by $G$. 
In our previous paper\cite{Tamaki}, we investigated two types of 
non-Abelian black hole
solutions in BD theory (Proca black hole and Skyrme black hole)
and found that there are several advantages to discussion in  the Einstein 
conformal frame rather than in the BD frame. 
For example, the definition of thermodynamical variables of the solutions
and the stability analysis using catastrophe theory.
Although, of course, we can reach the same results in the BD
frame,  we  transform the Einstein frame\cite{Dicke}
as a matter of convenience. The conformal transformation in the present model 
is written by
%%%%%%%%%%%%%%
\begin{eqnarray}
\hat{g}_{ab}=\frac{\phi}{\phi_{0}  } g_{ab}.
\label{B2}
\end{eqnarray}
%%%%%%%%%%%%%%
Then, we find the equivalent action $\hat{S}=S/\phi_{0}$ 
given as\cite{AD1} 
%%%%%%%%%%%%%%%%
\begin{equation}
\hat{S}  =  \displaystyle\int d^{4}x \sqrt{-\hat{g}}\left[
\frac{1}{2\kappa^{2}} \hat{R} -
\frac{1}{2}\hat{\nabla}_{\alpha}\varphi\hat{\nabla}^{\alpha}\varphi
+{1 \over \phi_{0}}\hat{L}_{m}\right],     \label{B3} 
\end{equation}
%%%%%%%%%%%%%%%
where
\begin{equation}
\varphi\equiv
\frac{1}{\kappa\beta}\ln \left({\phi \over
\phi_{0} }\right) , 
\end{equation}
%%%%%%%%%%%%%%%
\begin{equation}
\beta\equiv\left(\frac{2\omega+3}{2}
\right)^{-1/2}.
\label{B4}
\end{equation}
%%%%%%%%%%%%%%%
Variables with a caret  $\hat{}$  denote those in the
Einstein frame. $\hat{L}_{m}$ is different from $L_{m}$ unless the matter
field is conformally invariant, as with the Yang-Mills field.
The constant $\phi_{0}\equiv 2(2+\omega)/(2\omega +3)$ is the value 
of the BD scalar field $\phi$ 
at spatial infinity,  which guarantees that we can regard $G$ as 
Newton's gravitational 
constant\cite{BD}. 
In this paper we investigate the YMH system, 
where Lagrangian $\hat{L}_{m}$ is written as 
%%%%%%%%%%%%%%%
%%%%%%%%%%%%%%%
\begin{equation}
\hat{L}_{m}  =  -\frac{1}{4}F^{a}_{\;\mu\nu}F^{a\mu\nu}
-\frac{1}{2}e^{-\kappa \beta \varphi} 
(  D_{\mu}\Phi^{a}  )(  D^{\mu}\Phi^{a}  )
-\frac{\lambda}{4}e^{-2 \kappa \beta \varphi}
(\Phi^{a}\Phi^{a}-v^{2})^{2}
\label{B5}     
\end{equation}
%%%%%%%%%%%%%%%
$F^{a}_{\;\mu\nu}$ is the field strength of the SU(2) YM field 
and expressed by its potential $A^{a}_{\;\mu}$ as
%%%%%%%%%%%%%%%
\begin{equation}
F^{a}_{\;\mu\nu} =  \partial_{\mu}A^{a}_{\;\mu}-
\partial_{\nu}A^{a}_{\;\nu}+e \epsilon^{abc}A^{b}_{\;\mu}A^{c}_{\;\nu},
\label{B6}     
\end{equation}
%%%%%%%%%%%%%%%
with the gauge coupling constant $e$.
$\Phi^{a}$ is the 
real triplet Higgs field and $D_{\mu}$ is the covariant derivative: 
%%%%%%%%%%%%%%%
\begin{equation}
D_{\mu}\Phi^{a} =  \partial_{\mu} \Phi^{a}+
e \epsilon^{abc}A^{b}_{\;\mu}\Phi^{c}.
\label{B7}
\end{equation}
%%%%%%%%%%%%%%%
%%%%%%%%%%%%%%%
The theoretical parameters $v$ and $\lambda$ are the 
vacuum expectation value  and the self-coupling 
constant of the Higgs field, respectively. 
Below we consider them only in the Einstein conformal frame, 
and drop the caret $\hat{}$.

We assume that a space-time is static and spherically
symmetric, in which case the metric is written as
%%%%%%%%%%%%%%%%
\begin{equation}
ds^{2}=-\left[1-\frac{2Gm(r)}{r}\right]e^{-2\delta (r)}dt^{2}+
\left[1-\frac{2Gm(r)}{r}\right]^{-1}dr^{2}+r^{2}d\Omega^{2}.
\label{B8}
\end{equation}
%%%%%%%%%%%%%%%%
For the matter fields, we adopt the hedgehog ansatz given by 
%%%%%%%%%%%%%%%%
\begin{equation}
\Phi^{a}  =  v \mbox{\boldmath $r$}^{a} h(r), 
\label{B12} 
\end{equation}
%%%%%%%%%%%%%%%%
\begin{equation}
A^{a}_{\;0}=0, \label{B13}
\end{equation}
%%%%%%%%%%%%%%%%
\begin{equation}
A^{a}_{\;\mu}  =  \omega^{c}_{\;\mu}\epsilon^{acb} 
\mbox{\boldmath $r$}^{b}
\frac{1-w(r)}{er}, \;\;\;\;\; (\mu=1,2,3),
\end{equation}
%%%%%%%%%%%%%%%%
where $ \mbox{\boldmath $r$}^{a} $ and $\omega^{c}_{\;\mu}$ 
are a unit radial vector in the internal space and a triad, 
respectively. 

Variation of the action  (\ref{B3}) with the matter Lagrangian (\ref{B5})
leads to the field equations
%%%%%%%%%%%%%%%
\begin{equation}
\frac{d\delta}{d\tilde{r}} = -\frac{  8\pi\tilde{r}\tilde{v}^{2}\tilde{K}  }
{  \phi_{0}  },
\label{MPBD4-1} 
\end{equation}
%%%%%%%%%%%
\begin{equation}
\frac{d\tilde{m}}{d\tilde{r}}= 
\frac{  4\pi\tilde{r}^{2}\tilde{v}^{2}  }{  \phi_{0}  }
\left[
f\tilde{K}+\tilde{U}
\right],
\label{MPBD4-2}
\end{equation}
%%%%%%%%%%%
\begin{equation}
\frac{d^{2}w}{d\tilde{r}^{2}} =  
\frac{1}{f}
\left[
\frac{1}{2}\frac{  \partial \tilde{U}  }{  \partial w  }
+\frac{  8\pi\tilde{r} \tilde{v}^{2}\tilde{U}  }{  \phi_{0}  } 
\frac{dw}{d\tilde{r}} 
-\frac{  2\tilde{m}  }{  \tilde{r}^{2}  }\frac{dw}{d\tilde{r}}   
\right],
\label{MPBD4-3}
\end{equation}
%%%%%%%%%%%
\begin{equation}
\frac{d^{2}h}{d\tilde{r}^{2}} = \frac{dh}{d\tilde{r}}  
\left(  \beta\frac{ d\tilde{\varphi} }{d\tilde{r}}
-\frac{1}{\tilde{r}}  \right)
+\frac{1}{f}\left[  
e^{-\beta\tilde{\varphi} }  \frac{  \partial \tilde{U}  }{  \partial h  }
+\frac{8\pi\tilde{r} \tilde{v}^{2}\tilde{U}}{\phi_{0}}\frac{dh}{d\tilde{r}}
-\frac{1}{\tilde{r}}\frac{dh}{d\tilde{r}}  
\right],
\label{MPBD4-4}  
\end{equation}
%%%%%%%%%%%%%%%
\begin{equation}
\frac{d^{2}\tilde{\varphi}}{d\tilde{r}^{2}} = 
-\frac{2}{\tilde{r}}\frac{d\tilde{\varphi}}{d\tilde{r}}
-\frac{4\pi\beta\tilde{v}^2e^{-\beta\tilde{\varphi}  }}{\phi_{0}}
\left(  \frac{dh}{d\tilde{r}}  \right)^{2}
+\frac{1}{f}
\left[
\frac{  8\pi\tilde{v}^{2}  }{  \phi_{0}  }\left(
\frac{  \partial\tilde{U}  }{  \partial\tilde{\varphi}   }
+\tilde{r}\tilde{U}\frac{d\tilde{\varphi}}{d\tilde{r}}\right)
-\frac{2\tilde{m} }{\tilde{r}^{2}  }\frac{d\tilde{\varphi}  }{d\tilde{r}  } 
\right],
\label{MPBD4-5} 
\end{equation}
%%%%%%%%%%%%%%%%
where
%%%%%%%%%%%%%%%%%
%%%%%%%%%%%%%%%%%
\begin{equation}
f \equiv  1-\frac{2\tilde{m}}{\tilde{r}},
\end{equation}
%%%%%%%%%%%%%%%%%
\begin{equation}
\tilde{U} \equiv  
\frac{(1-w^{2})^{2}}{2\tilde{r}^{4}}
+e^{-\beta\tilde{\varphi}  }\left(  \frac{ wh }{ \tilde{r} }  \right)^{2}
+\frac{  \tilde{\lambda}  }{4}e^{-2\beta\tilde{\varphi} }(h^{2}-1)^{2},
         \label{MPpotential}
\end{equation}
%%%%%%%%%%%%%%%%%
\begin{equation}
\tilde{K} \equiv    
\frac{1}{\tilde{r}^{2}}\left(  \frac{dw}{d\tilde{r}}  \right)^{2}
+\frac{  e^{-\beta\tilde{\varphi}}  }{2}
\left(  \frac{dh}{d\tilde{r}}  \right)^{2}  
+\frac{\phi_{0}}{  16\pi \tilde{v}^{2}  }
\left( \frac{ d\tilde{\varphi} }{d\tilde{r}} \right)^{2}.     
\label{MPkinetic}     
\end{equation}
%%%%%%%%%%%%%%%
We have introduced the following
dimensionless variables:
%%%%%%%%
\begin{equation}
\tilde{r}=evr, \;\; \tilde{m}=Gevm, \;\; \tilde{\varphi}
=\kappa \varphi,
\label{B14} 
\end{equation}
%%%%%%%%
and dimensionless parameters:
\begin{equation}
\tilde{v}=v \sqrt{G}, \;\; \tilde{\lambda}=\lambda/e^{2}.
\label{B15}
\end{equation}
%%%%%%%%
Although the solution exists when $v\sim M_{Pl}$, where $M_{Pl}$ is
Planck mass, 
it can be described 
by a classical field configuration in the limit of a weak gauge coupling constant $e$,
because its Compton wavelength $\sim e/v$ is much smaller than the radius of the 
classical monopole solution $\sim 1/ev$ in this case. Moreover, since the energy density 
$\sim e^{2}v^{4} \ll M_{Pl}^{4}$, we can treat the system classically.

The boundary conditions at spatial infinity are
%%%%%%%%%%%%%%%%%%
\begin{equation}
\lim_{r\rightarrow \infty} m = M< \infty,  \label{Minf}
\end{equation}
%%%%%%%%%%%%%%%%%%
\begin{equation}
\lim_{r\rightarrow \infty}\delta =0,
\end{equation}
%%%%%%%%%%%%%%%%%%
\begin{equation}
\lim_{r\rightarrow \infty}\kappa\beta\varphi =
\lim_{r\rightarrow \infty}\ln \left(\frac{\phi}{\phi_{0}}\right)=0,  
\label{B9}
\end{equation}
%%%%%%%%%%%%%%%%%%
\begin{equation}
\lim_{r\rightarrow \infty}h=1,
\end{equation}
%%%%%%%%
\begin{equation}
\lim_{r\rightarrow \infty}w=0.    \label{B16}
\end{equation}
%%%%%%%%
These conditions imply that the space-time approaches a flat one and
that the solution is a charged object. The field equations are singular at the 
origin and at the
event horizon for a self-gravitating monopole and its black hole solution, respectively. 

For a self-gravitating monopole solution, we require regularity
at the origin. Expanding the Eqs.~(\ref{MPBD4-1})-(\ref{MPBD4-5}) 
around the origin,
we find that regularity is guaranteed when the fields behaves as
%%%%%%%%
\begin{equation}
w(\epsilon) \approx 1
+\frac12  C_w \epsilon^2+O(\epsilon^3),
\label{Bparticle1}  
\end{equation}
%%%%%%%%
\begin{equation}
h(\epsilon) \approx C_h \epsilon+O(\epsilon^3),
\label{Bparticle2}  
\end{equation}
%%%%%%%%
\begin{equation}
\tilde{\varphi}(\epsilon) \approx C_{\varphi}
-\frac{  2\pi \beta  }{  \phi_{0}  }e^{  -\beta C_{\varphi}  }\tilde{v}^{2}
\left(  C_{h}^{2}+\frac{ \tilde{\lambda} }{3}e^{  -\beta C_{\varphi}  }  \right)
\epsilon^2+O(\epsilon^3),
\label{Bparticle3}  
\end{equation}
%%%%%%%%
\begin{equation}
\delta(\epsilon) \approx C_{\delta} 
-\frac{ 4\pi\tilde{v}^{2}  }{  \phi_{0}  }\left(  C_{w}^{2}+
\frac{ e^{-\beta C_{\varphi}}  }{2}C_{h}^{2}  \right)
\epsilon^2+O(\epsilon^3),
\label{Boundary particle4}  
\end{equation}
%%%%%%%%
\begin{equation}
\tilde{m} (\epsilon) \approx
-\frac{ 2\pi\tilde{v}^{2}  }{ \phi_{0}  }\left(  
C_{w}^{2}+e^{  -\beta C_{\varphi}  }C_{h}^{2}+\frac{\tilde{\lambda}}{6}
e^{  -2\beta C_{\varphi}  }
\right) 
\epsilon^3+O(\epsilon^4).
\label{Bparticle5}  
\end{equation}
%%%%%%%%%%%%%%%%%%
$C_w$, $C_h$ and $C_{\varphi}$ are constant, which should be
determined iteratively so that 
the boundary conditions (\ref{B9})-(\ref{B16}) 
are satisfied. We also require that no singularity exists, i.e., 
%%%%%%%%%%%%%%%%%%
\begin{eqnarray}
m(r)<\frac{r}{2G}\   . \label{Bparticle11}
\end{eqnarray}
%%%%%%%%%%%%%%%%%%

In the case of the monopole black hole, we assume
the existence of a regular event horizon at $r=r_H$. For the metric
%%%%%%%%%%%%%%%%%%
\begin{equation}
m_{H}\equiv m(r_H)=\frac{r_{H}}{2G},  \label{MHbound}
\end{equation}
%%%%%%%%%%%%%%%%%%
\begin{equation}
\delta_{H}\equiv\delta(r_H)<\infty.
\end{equation}
%%%%%%%%%%%%%%%%%%
For the matter fields, the square brackets in 
Eqs.~(\ref{MPBD4-3})-(\ref{MPBD4-5}) 
must vanish at the horizon. Hence we find that 
%%%%%%%%%%%%%%%%%
%%%%%%%%%%%%%%%%%
\begin{equation}
\left. \frac{dw}{d\tilde{r}} \right|_{ \tilde{r}=\tilde{r}_{H} }
=\frac{w_{H}\phi_{0}}{F}
\left(1-w^{2}_{H}-h^{2}_{H}\tilde{r}^{2}_{H}
e^{   -\beta\tilde{\varphi}_{H}  }  \right)       
\label{MPBD4-6}   
\end{equation}
%%%%%%%%%%%%%%%%%
\begin{equation}
\left.\frac{dh}{d\tilde{r}}\right|_{  \tilde{r}=\tilde{r}_{H}  }
=-\frac{h_{H}\phi_{0}}{F}
\left[2w^{2}_{H}+\tilde{\lambda}\tilde{r}^{2}_{H}
(h^{2}_{H}-1 )e^{   -\beta\tilde{\varphi}_{H} } \right] 
\label{MPBD4-7}    
\end{equation}
%%%%%%%%%%%%%%%%%
\begin{equation}
\left.\frac{d\varphi}{d\tilde{r}}\right|_{\tilde{r}=\tilde{r}_{H}}
=\frac{  2\pi\beta\tilde{v}^{2}e^{-\beta\tilde{\varphi}_{H}}  } {  F  }
\left[   \tilde{\lambda}\tilde{r}^{2}_{H}e^{   -\beta\tilde{\varphi}_{H}   }
(h^{2}_{H}-1)^{2}+2w^{2}_{H}h^{2}_{H} \right],               
\label{MPBD4-8}
\end{equation}
%%%%%%%%%%%%%%%
where
%%%%%%%%
\begin{eqnarray}
F&=& 2\pi\tilde{v}^{2}\tilde{r}_{H}\left[2\tilde{r}^{-2}_{H}(
    1-w^{2}_{H}    )^{2}+4e^{   -\beta\tilde{\varphi}_{h}   }
w^{2}_{H}h^{2}_{H}+\tilde{\lambda}\tilde{r}^{2}_{H}
(    h^{2}_{H}-1   )^{2}e^{   -2\beta\tilde{\varphi}_{H} }\right]
 -\phi_{0}\tilde{r}_{H}  
\label{bunbo}  .
\end{eqnarray}
%%%%%%%% 
We also require that no singularity exists outside the
horizon, i.e., 
%%%%%%%%%%%%%%%%%%
\begin{eqnarray}
m(r)<\frac{r}{2G}\  ~~~~~{\rm  for} ~~~ r>r_{H}\ . \label{B11}
\end{eqnarray}
%%%%%%%%%%%%%%%%%%

Hence, we should determine the values of $C_{w}$, $C_{h}$  and $C_{\varphi}$ 
(self-gravitating monopole case) or $\tilde{\varphi}_{H}$, $w_{H}$,  and $h_{H}$ 
(monopole black hole case) iteratively so that the boundary conditions 
at infinity are fulfilled. In this sense these are shooting parameters.
It is a remarkable property of the 
nontrivial structure of many systems discussed before  
that we can not choose arbitrarily the values both of the fields
and of their derivatives at the horizon.
When we investigate the internal structure
of  non-Abelian black holes, we may expect the existence of the 
inner (Cauchy) horizon. There, the fields must satisfy the same type
of constraint such as Eqs.~(\ref{MPBD4-6})-(\ref{MPBD4-8}). However, the 
shooting parameters on the black hole horizon which satisfy the 
boundary condition at infinity do not necessarily fulfill the 
constraint at the inner horizon. By this behavior the mass function 
diverges and a mass inflation phenomenon is realized inside of the black hole
event horizon for non-Abelian black holes \cite{Donets}.

A non-trivial solution does not necessarily exist for the given theoretical
parameters. However, for arbitrary values of 
$\tilde{v}$ and $\tilde{\lambda}$, there exists a RN black hole solution such as 
%%%%%%%%
\begin{eqnarray}
w= 0 ,\ h = 1 ,\ \delta = 0,\ 
\phi = \phi_{0},\ \tilde{m}(\tilde{r})= \tilde{M}-
\frac{2\pi\tilde{v}^{2}}{\phi_{0}\tilde{r}}\ .
\label{RN1} 
\end{eqnarray}
%%%%%%%% 
$\tilde{M}$ is the gravitational mass at spatial infinity and
$\tilde{Q} \equiv 2\tilde{v}\sqrt{\pi/\phi_{0}}$ is the magnetic charge of the
black hole. The radius of the event horizon of the
RN black hole is constrained to be $\tilde{r}_H \geq \tilde{Q}$.
The equal sign case implies an extreme solution. 
If we equate $\beta=0$, (i.e., $\omega \to \infty$) the 
model recovers the EYMH system and we find a self-gravitating monopole and its  
black hole solution investigated in Ref.~\cite{Lee,Ortiz,BFM,Aichelburg,Tachi}.
In addition, if we put $G=0$, i.e., neglect the gravitational effect,
the model recovers the YMH system, which has a famous 't~Hooft-Polyakov
solution for a regular magnetic monopole\cite{'t Hooft}. Further,
assuming $\lambda=0$, we obtain the BPS monopole solution\cite{BPS}, 
%%%%%%%%
\begin{eqnarray}
w(\tilde{r})=\frac{  \tilde{r}  }{  \sinh \tilde{r}  },\ \ 
h(\tilde{r})=\coth \tilde{r}-\frac{ 1 }{ \tilde{r} }.  
\label{BPSlimit} 
\end{eqnarray}
%%%%%%%% 

In the next section, we show the solutions 
of a self-gravitating monopole both in GR and  BD theory.

%%%%%%%%%%%%%%%%%%%%%%%%%%%%%%%%%%%%%%%%%%%%%%%%%%
%%%%%%%%%%
%                                                          %
\section{Self-gravitating monopole in Brans-Dicke theory}    %
%                                                          %
%%%%%%%%%%%%%%%%%%%%%%%%%%%%%%%%%%%%%%%%%%%%%%%%%%
%%%%%%%%%%
The self-gravitating monopole  in
GR were investigated in detail by several authors
\cite{Lee,Ortiz,BFM,Aichelburg,Tachi} and found to have
many interesting properties, some
of which can not be seen for the monopole solution in  flat space-time.
We  list their properties. 
(1)~The  distribution of the non-Abelian structure decays exponentially 
with respect to the  
distance from the origin and the monopole has its core radius $\sim 1/ev$.
(2)~The solutions are characterized by the node number of the
YM potential, which is different the from topological winding number of the monopole. 
(3)~There exists the maximum parameter $v_{max}$ beyond
which there is no monopole solution. $v_{max}$ weakly depends on 
$\lambda$.
(4)~There exists the critical parameter 
$v_{extreme}$ for the small $\tilde{\lambda}$. 
At $v=v_{extreme}$, the monopole solution changes to extreme RN
solution.
Between $v_{extreme}$ and $v_{max}$,
two monopole solutions with a different mass appear for each value of $v$. 
The higher mass solution is unstable while the 
lower  one is stable. $v_{extreme}$ also weakly
depends on $\lambda$.

Now we turn to the monopole solutions in BD theory and compare their
properties with those in GR from the above points of view. 
Solving the system 
(\ref{MPBD4-1})-(\ref{MPkinetic}) 
with the boundary conditions (\ref{Minf})-(\ref{Bparticle11}),
we obtain gravitating monopole solutions.
Although they
are found only numerically, there exists a rigorous 
proof of the existence of self-gravitating monopole and its black 
hole solution in the $\tilde{\lambda}\to \infty$ case 
 in GR \cite{BFM}. We expect
that the same discussion holds in the more general case including BD theory.

We show the field configurations of the gravitating monopole in Figs.~1.
We chose the parameters as $\tilde{v}=0.1$, $\tilde{\lambda}=0.1$ 
and several values of $\omega = -1.4$, $-1.0$, $0$, $\infty$.
The $\omega =\infty$ case corresponds to GR.
(a) is $\tilde{r}$-$h,w$ diagram. 
We can see that the configurations of the YM potential and the Higgs field
hardly depend on the BD parameter $\omega$. This implies that the 
solution can have almost the same structure as the 't Hooft-Polyakov monopole
in flat space-time, where the repulsive force of the gauge field balances
with the attractive force of the Higgs field. (b) and (c) are 
$\tilde{r}$-$\tilde{m}_g$ and $\tilde{r}$-$\delta$ diagram, respectively.
Since $\tilde{m}_g$ is defined as
%%%%%
\begin{equation}
\tilde{m}_g \equiv \tilde{m} + \frac{2\pi\tilde{v}^2}{\phi_0\tilde{r}},
\end{equation}
%%%%%
it diverges at the origin.
The mass function $\tilde{m}_g$ strongly depends on $\omega$. 
For a different BD parameter
each solution has a different effective global gauge charge $\tilde{Q}$.
For the small $\omega$, the charge is small, i.e.,
the monopole in  BD theory has a smaller charge than that in GR.
Since there is a prefactor $\tilde{Q}$ in the
$\delta$ and $\tilde{m}$ equations, it
contributes directly to the behavior
of the metric functions. There is another factor which may change the
behavior of metric functions. It is the BD scalar field $\varphi$.
However from Fig. 1~(d), $\tilde{\varphi}\lsim0.07$ even for
$\omega =-1.4$, which  implies the $e^{-\beta \tilde{\varphi}}\approx 1$. 
$d \tilde{\varphi} / d\tilde{r}$ term is also smaller than other 
terms in the field equations.
Hence we can neglect the $\tilde{\varphi}$ contribution to other fields in the 
{\it zeroth} approximation. However, we may not neglect the $\tilde{\varphi}$
contribution for the parameter range $v \approx v_{extreme}$
discussed later, where the gravitational effect plays an important roll.

In Fig.~2 we show  the distribution of the energy density 
$\rho\equiv -T^{0}_{\;0}$
for the same parameters as Fig.~1 (It is normalized as 
$\tilde{\rho}=\rho  G  /  (ev)^{2}  $). An arrow shows the core radius of the monopole.
The prefactor $1/\phi_0$ for the matter Lagrangian can be absorbed
by defining  the new variables and the effective
parameters as $\bar{A}^{a}_{\mu}=A^{a}_{\mu}/\sqrt{\phi_0}$,
$\bar{\Phi}=\Phi/\sqrt{\phi_0}$, $\bar{e}=\sqrt{\phi_0}e$,
$\bar{v}=v/\sqrt{\phi_0}$ and $\bar{\lambda}=\phi_0\lambda$.
Hence the core radius $\sim 1/\bar{e}\bar{v}=1/ev$ hardly
depends on $\phi_0$, i.e. BD parameter $\omega$.
The energy density is almost constant inside the monopole core, but 
it decays as $1/\tilde{r}^4$ outside. Note that although $w$
decays exponentially, the YM potential decays $\sim 1/r$ for a large
radius.
 
As for the monopole  in flat space-time, the uniqueness of the solution was
proven \cite{Maison}, which means there is no excited mode of 
this solution. As for gravitating counterpart, however, 
there exist radial excited 
solutions characterized by the
node number $n$ of the YM potential
for small $\tilde{v}$ in GR\cite{BFM}.  
We also find such excited solutions in BD theory.
In the $v \to 0$ limit, the Higgs field becomes constant and
the YM field recovers its apparent gauge symmetry. Then
the YM field is conformally invariant and the BD scalar field becomes
trivial by the regularity. As a result, the $n=1$ excited solution 
approaches a regular Bartnik-McKinnon solution \cite{BM} in the EYM system in this
limit as discussed in Ref.~\cite{BFM}.
Hereafter, we focus on the solution with $n=0$, i.e.,
the ground state solution.

%======

In GR, there is  the maximum value of
the vacuum expectation value $\tilde{v}_{max}$  for a fixed 
$\tilde{\lambda}$ where a regular 
monopole solution can exist \cite{BFM}. 
We can understand this intuitively as follows. We compare 
the mass of the monopole $\sim 4\pi v/e$ with its core radius $\sim 1/ev$. 
When $v$ becomes large, the mass of the monopole increases, as does the gravitational 
radius $\sim 8 \pi G v/e$, while its core radius shrinks. 
Hence, its core radius eventually  gets into gravitational radius
and the  solution may develop into the RN black hole being
swallowed in its non-trivial structure by the horizon. 
The existence of the boundary of the parameter is interesting also in the
context of the inflation. 
Since there is no non-trivial static configuration beyond the maximum parameter,
the solution must take a configuration of
the static RN black hole or a nonstatic configuration. One of the possibilities 
in the latter case is the inflating solution. Actually it is confirmed that
the boundary of the static solution  almost coincides with the  
parameter region where the monopole inflation occurs \cite{Vilenkin}.

In BD theory, it was shown that when we consider the global monopole, any 
inflating monopole eventually shrinks contrary to the case in GR\cite{Sakai}. 
Although we consider gauge monopole here, the BD scalar field may 
give a serious effect. Moreover it is found that
the gravitating monopole solution
in $N=4$ supersymmetric low-energy superstring theory  exists 
for arbitrary values of vacuum expectation value of the Higgs field, 
where the dilatonic field plays an important roll \cite{Harvey}. 
Since the BD scalar field can be regarded as the dilaton field in a
special coupling constant case as can be seen in Eq.~(\ref{B3}),
there is a possibility that  $\tilde{v}_{max}$ disappears.
Actually,   taking $\tilde{v}$ larger, we find the BD scalar field gives a larger 
contribution.  
Hence we are interested in examining the maximum value 
$\tilde{v}_{max}$ in BD theory.

First, we discuss analytically  the parameter region where the nontrivial
solutions exist following the analysis in Ref.~\cite{Lee}. 
For the monopole solution, the mass function behaves as Eq.~(\ref{Bparticle5})
around the origin by regularity.
At spatial infinity, the solution 
approaches  RN solution, 
%%%%%%%%%%%%%%%%
\begin{eqnarray}
\tilde{m}(\tilde{r})=\tilde{M}-\frac{2\pi \tilde{v}^{2}}{\phi_{0}\tilde{r}}
+\cdots\ .
 \label{mass-infty}
\end{eqnarray}
%%%%%%%%%%%%%%%%
From this boundary condition, $g^{rr}$ 
should have a minimum which 
corresponds to the maximum 
of the $\tilde{m}/\tilde{r}$. This occurs around 
$\tilde{r}_{m}\sim 4\pi\tilde{v}^{2}/\phi_{0}\tilde{M}$. 
For the monopole in flat space-time, we have 
%%%%%%%%%%%%%%%%
\begin{eqnarray}
\tilde{M}_{flat}= \frac{  4\pi\tilde{v}^{2}  }{ \phi_{0} } g(\tilde{\lambda})\ , 
 \label{flat-monopole}
\end{eqnarray}
%%%%%%%%%%%%%%%%
$g(\tilde{\lambda})$ takes values from $1$ to $1.787$ when 
$\tilde{\lambda}$ changes from $0$ to $\infty$.  
We assume  $\tilde{M}\propto \tilde{v}^{2}$ 
even for the gravitating monopole in both theories.
Under this assumption, $\tilde{r}_{m}\sim 1/g \sim 1$. 
So $g^{rr}$ takes a minimum 
value of about  $g^{rr}\sim 1-4\pi \tilde{v}^{2}/\phi_{0}$. 
When $\tilde{v}=\tilde{v}_{extreme}=\sqrt{\phi_{0}/4\pi}$, 
$g^{rr}$ takes $0$ and the degenerate horizon appears at $r=r_{extreme}$. 
Because the region $\tilde{r}<\tilde{r}_{extreme}$ is separated infinitely 
in geodesic 
distance from the outer region\cite{CosmicString}, 
the monopole structure could be thought 
to be confined in the region 
$\tilde{r}<\tilde{r}_{extreme}$ and the solution looks like a RN black hole
with the horizon radius $r_{extreme}$. Naively 
$\tilde{v}_{extreme}$ gives the 
maximum value of $\tilde{v}$.
As a result, we can expect that the critical value $\tilde{v}_{extreme}$ becomes large
for the large $\phi_0$, i.e., the small value of $\omega$. 

Breitenlohner et al. confirmed the above discussion numerically in
GR ($\phi_0=1$) and found
the maximum value $\tilde{v}_{max}$ for the  fixed $\tilde{\lambda}$
by examining the minimum of $g^{rr}$ \cite{BFM}. 
Moreover they found the parameter region of $\tilde{\lambda}$ where 
$\tilde{v}_{extreme}$ is not the maximum value of $\tilde{v}$.
In the parameter region $\tilde{v}_{extreme}<\tilde{v}<\tilde{v}_{max}$,
there are two different nontrivial solutions with 
the same $\tilde{v}$ which is not obvious in the above simple analysis. 
On $\tilde{v}$-$\tilde{M}$ diagram, these solutions construct two solution 
branches divided by a cusp structure at $\tilde{v}=\tilde{v}_{max}$.
The lower mass branch ends up with $\tilde{v}_{max}$ and the
higher mass branch is connected to the RN black hole branch
at $\tilde{v}=\tilde{v}_{extreme}$. 
The cusp structure is important for the stability analysis by
using catastrophe theory\cite{Tachi}.
For the larger $\tilde{\lambda}$,
$\tilde{v}_{max}$ and $\tilde{v}_{crit}$ coincide\cite{lambdabig}.
Since $\tilde{v}_{max}$ takes a value very close to $\tilde{v}_{extreme}$,
we search $\tilde{v}_{extreme}$ by the behavior of $g^{rr}$ first and
estimate $\tilde{v}_ {max}$.

In the same manner, we examine $\tilde{v}_{max}$ in BD theory with $\omega=0$.
Figs.~3 shows the field configurations of the
monopole with $\tilde{\lambda}=0.2$ and around the maximum parameter
$\tilde{v}_{max}$ ((a) $\tilde{r}$-$g^{rr}$ (b) $\tilde{r}$-$h$, $w$). 
We can find that $g^{rr}$ has the  minimum
around $\tilde{r} = \tilde{r}_{extreme} \sim 1.53$ and the minimum value 
decreases as $\tilde{v}$ becomes larger. 
When  $\tilde{v}$ approaches 
$\tilde{v}_{max}$, the behavior of the YM potential and the Higgs field 
changes more rapidly near the $\tilde{r}_{extreme}$ and 
takes $w\approx 0$ and $h\approx1$, i.e., the
solution can be approximated by a RN black hole solution outside of 
$\tilde{r}_{extreme}$.
When we have $\tilde{v}_{extreme}$, 
the horizon degenerates and the outer region  becomes the extreme RN
black hole. Since the gravitational effect becomes strong around this parameter region as seen from Fig. 3 (a), 
the amplitude of $\tilde{\varphi}$ and the dependence of $w$ and $h$ on the BD parameter 
can not be negligible. 
We show the $\tilde{v}_{max}$-$\tilde{\lambda}$ diagram with 
$\omega=0$ and $\infty$  in Fig.~4 (the solid lines). 
Though $\tilde{v}_{max}$ is larger compared to the estimation 
$\tilde{v}_{extreme}=\sqrt{\phi_{0}/4\pi}\sim 0.282$ (for GR), 
$\sim 0.325$ (for $\omega =0$) due to the naive estimation, 
we can obtain the relevant dependence of  $\tilde{v}_{max}$ on $\phi_{0}$ 
(or $\omega$). If we take a more accurate value of $r_m$ calculated 
numerically, we have the better estimation of $\tilde{v}_{extreme}\sim 0.369$ 
(for GR) and $\tilde{v}_{extreme}\sim 0.497$ 
(for $\omega =0$)
As we describe later, the similar 
phenomena can be seen in monopole black hole. These are also shown in this diagram 
by the dotted lines (In this diagram, $\tilde{r}_{H}=0.4$.).

%%%%%%%%%%%%%%%%%%%%%%%%%%%%%%%%%%%%%%%%%%%%%%%%%%
%%%%%%%%%%
%                                                          %
\section{Monopole black hole in Brans-Dicke theory}    %
%                                                          %
%%%%%%%%%%%%%%%%%%%%%%%%%%%%%%%%%%%%%%%%%%%%%%%%%%
%%%%%%%%%%

The monopole black hole solution in GR has the following 
properties in addition to (1)-(2) of Sec.III.
The number of the  solutions depends on the parameters 
complicatedly, since another parameter, the horizon radius,
is added in the black hole case. We show the 
schematic diagram of $\tilde{v}$-$\tilde{r}_H$  
for small $\tilde{\lambda}$ in Fig.~5 \cite{note1}.
In the shadowed region, the nontrivial solution exists.
The $\tilde{r}_H=0$ axis corresponds to the regular monopole
solution. 
(5) For a fixed value of $\tilde{r}_H<\tilde{r}_H^{\ast}$ (See Fig. 5), we find two
characteristic values $\tilde{v}_{max}$ and $\tilde{v}_{extreme}$.
They are extensions of properties (3) and (4) in the regular monopole case discussed in 
the previous section. In the region (i) in Fig. 5
($\tilde{v}_{extreme}<\tilde{v}<\tilde{v}_{max}$), there are two nontrivial
solutions with different mass. The solution branch with higher mass is
connected to the extreme RN black hole solution in the $\tilde{v}$-$\tilde{M}$
diagram although the horizon radius changes  discontinuously.
(6) The solution in the lower mass branch is stable while
the higher one is unstable. 
(7) For a  fixed value of $\tilde{v}<\tilde{v}^{\ast}$ (See Fig. 5), 
there are two characteristic
values $\tilde{r}_{H,max}$ and $\tilde{r}_{H,RN}$. In the region (ii) in 
Fig. 5
($\tilde{r}_{H,RN}<\tilde{r}<\tilde{r}_{H,max}$), there are two 
nontrivial solutions.
The solution branch with lower entropy (smaller $\tilde{r}_{H}$) is 
connected to the RN black hole at $\tilde{r}_{H}=\tilde{r}_{H,RN}$.
(8) The solution in the high entropy
branch is stable while the lower one is unstable.
(9) The regions (i) and (ii) disappear for $\tilde{\lambda}>\tilde{\lambda}_{crit}^{(i)}\sim 0.287$ and
$\tilde{\lambda}>\tilde{\lambda}_{crit}^{(ii)}\sim 0.696$, respectively.
(10) When we consider the thermodynamics of the
monopole black hole, its temperature diverges in the minimum mass limit
unlike the RN black hole. Through a evaporation process from a large
RN black hole, the solution experiences the first and/or second
order phase transitions.

%-configuration (r_h dependence)
We turn to the monopole black hole  in BD theory.
Solving the system 
(\ref{MPBD4-1})-(\ref{MPkinetic}) numerically
with the boundary conditions (\ref{Minf})-(\ref{B16}) and (\ref{MHbound})-(\ref{B11}),
we obtain a black hole solution.
Figs.~6 are the field configurations with the
parameters $\tilde{\lambda}=0.1$, $\tilde{v}=0.1$, $\omega=0$
and the horizon radius $\tilde{r}_{H}=0.2,\ 0.4$ and $0.6$
((a) $\tilde{r}$-$h$,$w$ (b) $\tilde{r}$-$\tilde{m}_{g}$ (c) $\tilde{r}$-$\delta$ 
(d) $\tilde{r}$-$\tilde{\varphi}$)
We also plot the gravitating monopole solution, i.e., $\tilde{r}_{H}=0$, for comparison.
As the horizon radius becomes large, the curvature of the space-time
outside the horizon would become small resulting in the variation of the
BD scalar field being small. Hence the difference between the solutions
in BD theory and in GR appears  most conspicuously in the monopole solution case. The boundary value of 
$w$ and $h$ approaches
$w(r_H) \to 0$ and $h(r_H) \to 1$ for a large horizon radius. Finally the solution
becomes a RN black hole in the $\tilde{r}_{H}=\tilde{r}_{H,RN}$ limit.
The behaviors of the solutions around this limit depend on the parameters
complicatedly as discussed later.

%-M-r_h diagram
In Fig.7(a), we show the $\tilde{M}$-$\tilde{r}_{H}$ diagram for
$\tilde{\lambda}=0.1$, $\tilde{v}=0.1$
and $\omega=-1.4$, $-1$, $0$ and $\infty$.
Dotted lines show RN solution branches. Note that the RN solution
in BD theory is different from that in GR because it has a different 
effective gauge
charge by the  factor $\phi_{0}$. Unlike the
RN black hole, the monopole black hole does not have a extreme limit
but has the $\tilde{r}_{H}=0$ limit, which 
corresponds to the gravitating monopole solution.
Fig.7(b) is a magnification around the maximum horizon radius of each theory.
As is the case of GR, there are two solutions in some ranges of horizon radius. 
This parameter region corresponds to the region (ii) in Fig.~5.
We call the upper (lower) branch a high (low) entropy branch.
In both theories, these branches form a cusp structure at the point $A$, where
the solution has the maximum horizon radius $\tilde{r}_{H,max}$. 
At the point $B$, the solution coincides with the RN  black hole solution
with the horizon radius $\tilde{r}_{H,RN}$. 
The solutions have a qualitatively similar structure in both theories. 
The maximum mass and the maximum horizon radius 
decrease when $\omega$
becomes small. Furthermore we find that the range of horizon radius where 
the low entropy branch exists becomes larger 
when we take the $\omega$ as small.
These behaviors are mainly due to the effect of the boundary value of $\phi_0$.
As we mentioned, $\bar{v}=\tilde{v}/\sqrt{\phi_0}$ 
can be considered as an effective
vacuum expectation value in BD theory. Since it becomes small
when $\omega$ becomes small, we find that the width of region (ii) in 
Fig. 5 becomes large.

%-high and low entropy branch
Next we consider a difference of  solution in each branch.
We consider the solution  with $\tilde{v}=0.1$,
$\tilde{\lambda}=0.1$ and $\tilde{r}_{h}=0.59$, and change the
BD parameter $\omega$. Figs.~8 and 9 show   configurations
of the fields ((a) $\tilde{r}$-$h$, $w$ (b) $\tilde{r}$-$\tilde{m}_{g}$
(c) $\tilde{r}$-$\tilde{\varphi}$ ).
In the high entropy branch (Fig.~9), the YM potential and the Higgs
field hardly depend on $\omega$. This is the same as the self-gravitating
monopole case and the structure is supported by the balance 
between these fields. On the other hand, they have
small amplitude in the low entropy branch (Fig.~10). The Higgs
field can be approximated as $h\approx 1$, which means that it
takes almost its vacuum value, while 
$w \approx 0$, where the YM potential decays as $\tilde{r}^{-1}$.
Hence,  what attracts the YM field is not the Higgs field
but the self-gravitational attractive force. Actually, 
the counterpart of the low entropy branch does not exist in  flat 
space-time. The dependence on $\omega$ in the low entropy
branch is due to the difference of the effective gauge charge again. The
horizon radius $\tilde{r}_{h}=0.59$ line crosses the low entropy
branch relatively near the point A (See Fig.8(b)). In the large $\omega$ (GR), 
however, it crosses around the point B, where the solution 
coincides with the RN black hole. Hence the field configurations approach
a trivial solution.

%-scalar mass
In the present model, the YM field obtains the mass through
a spontaneous symmetry breaking mechanism and the trace part
of the energy-momentum tenser does not vanish. In this case the BD 
scalar field  takes a nontrivial configuration and we can define 
the scalar mass
$M_s$ by the asymptotic behavior of the BD scalar field as 
%%%%%%%%%%%%%%%%
\begin{eqnarray}
\phi \sim \phi_{0}\left(  1+\frac{2GM_{s}}{r}  \right). 
\label{scalar charge}
\end{eqnarray}
%%%%%%%%%%%%%%%%
Fig.~10 shows $\omega$ dependence of the gravitational mass $M$ 
and scalar mass $M_s$ for fixed horizon radius $\tilde{r}_{h}=0.59$.
As the solution can be approximated by the RN black hole in the 
zeroth approximation, $M$ behaves as
%%%%%%%%%%%%%%%%
\begin{eqnarray}
\tilde{M} \approx \frac{1}{2\tilde{r}_h}
\left[\tilde{r}_h^2+4\pi \tilde{v}^2
-\frac{2\pi\tilde{v}^2}{ \omega+2}   \right]. 
\end{eqnarray}
%%%%%%%%%%%%%%%%
This gives good agreement with the $\omega$ dependence in Fig.~10(a).
$\tilde{M}_s$ can be expressed as \cite{Will}

%%%%%%%%%%%%%%%%
\begin{eqnarray}
\tilde{M}_{s}=\frac{1-2s}{2\omega +3}\tilde{M} ,
\label{sense}
\end{eqnarray}
%%%%%%%%%%%%%%%%
where $s$ is the sensitivity of a self-gravitating object. Though $s$ varies 
as  $\omega$ changes, its configuration is negligible
for large $\omega$ and $\tilde{M}_s$ behaves as $\sim 1/\omega$.
Since $s \to 1/2$ in the $\omega \to -3/2$ limit, the fraction in 
Eq.~(\ref{sense}) is indeterminate. However, 
Fig. 11(b) shows that $\tilde{M}_s$ takes a non-zero finite value.
As $\tilde{M}$ takes almost the same value in each branch, the solution in the
high entropy branch has smaller sensitivity than the lower branch.

%-\lambda_critical
In GR, it is considered that when $\lambda$ becomes large,
the parameter point where the $\tilde{r}_{H,RN}=\tilde{r}_{H,max}$
in Fig.~5s shifts to the left side, i.e., smaller $\tilde{v}$ and the region (ii) becomes
small. When $\tilde{\lambda}$ takes the critical value 
$\tilde{\lambda}_{crit}^{(ii)}$, the region (ii)  finally 
disappears \cite{BFM,Torii}.
Although we do not examine the definite value of 
$\tilde{\lambda}_{crit}^{(ii)}$ in BD theory,
the dependence on $\omega$ would be small because 
$\tilde{\lambda}=\lambda/e^2=\bar{\lambda}/\bar{e}^2$
does not depend on $\phi_0$.

%-Stability
The stability of these exotic objects is one of the main interests. 
Aichelburg \cite{Aichelburg} analyzed the monopole black hole and its stability 
via the linear perturbation method in GR and showed that most of  the 
solutions are stable within the limit of $\tilde{\lambda}\rightarrow \infty$.
They also obtained the solutions in the finite $\tilde{\lambda}$ case
and suggested that if we have two monopole black hole 
solutions for a fixed value of $\tilde{v}$, the low energy one is stable 
while the other is not. Similar works are  found in \cite{BFM},  
where the authors considered the weak coupling limit. 
We examined stabilities via catastrophe theory\cite{Tachi} 
and obtained a unified picture of the stabilities of the gravitating
monopole and the monopole black hole. 
In catastrophe theory\cite{CAT}, we discuss the shape of a potential function 
as the variation of the control parameters. 
In that paper, we adopted $\tilde{\lambda}$, $\tilde{v}$ and $\tilde{M}$
as the control parameters, entropy $S$ as the potential function and
$\delta(r_h)$ as the state variable, and found that the
system is classified into a swallow tail catastrophe.
In general, a swallow tail catastrophe has three control parameters 
$a$, $b$, $c$ and 
one state variable $x$ and its potential function is described as 
%%%%%%%%
\begin{eqnarray}
V=\frac{1}{5}x^{5}+\frac{1}{3}ax^{3}+\frac{1}{2}bx^{2}+cx  \ .
\label{Swallow} 
\end{eqnarray}
%%%%%%%% 
We regard the parameters as 
%%%%%%%%
\begin{eqnarray}
a=a(\tilde{\lambda},\tilde{v}),\ \ b=b(\tilde{M},\tilde{v}) ,\ \ c=0.
\label{Control} 
\end{eqnarray}
%%%%%%%% 
$\tilde{\lambda}^{(ii)}_{crit}$ is obtained by 
$a(\tilde{\lambda}^{(ii)}_{crit}, 0)=0$. 
Although we have a new control parameter
$\omega$ in BD theory, 
the situation does not change. By analyzing the  dependence
of the solution on the parameters, Eq.~(\ref{Control}) is  changed to 
%%%%%%%%
\begin{eqnarray}
a=a(\tilde{\lambda},\tilde{v},\omega),\ \ 
b=b(\tilde{M},\tilde{v},\omega), \ \ c=0,
\label{Control2} 
\end{eqnarray}
%%%%%%%% 
i.e., the BD parameter is not an intrinsic parameter but a  dummy 
parameter in catastrophe theory, and
does not result in qualitative difference from GR. 
This is the same as in the neutral type non-Abelian black hole
case\cite{Tamaki}. As a result, the
monopole black hole in BD theory is also classified as a swallow tail
catastrophe. From this we find that the solution in the high entropy 
branch and RN black hole with larger horizon radius than 
$\tilde{r}_{H,RN}$ are stable, while the solutions in the low entropy 
branch and RN black hole with smaller horizon radius than 
$\tilde{r}_{H,RN}$ are unstable.
 
%- uniqueness and no-hair
These properties are interesting in the context of 
black hole no-hair conjecture and the uniqueness of the solution. 
For $\tilde{\lambda}>\tilde{\lambda}^{(ii)}_{crit}$, there are two
solutions (monopole black hole and RN black hole) when we fix the 
global charge $M$ and $Q$. This is the
counterexample of the strong no-hair conjecture, i.e., if the global 
charges are fixed, there exists only one solution.
Moreover when $\tilde{\lambda}$ is smaller than 
$\tilde{\lambda}^{(ii)}_{crit}$,
there is a mass range where two stable solutions (monopole black hole in 
the high entropy branch and RN black hole)
exist for a fixed mass. This violates even the weak no-hair
conjecture which states that if the global 
charges are fixed, there exists only one {\it stable} solution. 
These behaviors are shared by both theories.
In GR, however, we have no idea how to identify whether a black hole is 
a monopole black hole or RN black hole from infinity, because
the NA structure and the Higgs field decrease exponentially. 
On the other hand, in BD theory
we can obtain the information about the scalar mass, which
can be used to tell the type of  black holes.
Hence we can identify each black hole uniquely from infinity.

%-Temperature
When we take into account the quantum effect, a black hole loses
its energy by the Hawking evaporation process. The evolution of the
monopole black hole by this process is rather complicated.
First, we show an inverse temperature in the $\tilde{\lambda}=1.0$,
$\tilde{v}=0.1$ and  $\omega=0$, $\infty$ (GR) in Fig.~11(a). 
The inverse temperature of the RN black hole has the minimum 
$1/T=6\sqrt{3}\pi Q$ at $\tilde{M}=2\tilde{Q}/\sqrt{3}$ and diverges at
the extreme limit $\tilde{M}=\tilde{Q}$, where the evaporation ends. 
On the other hand, the inverse temperature of the monopole black hole
becomes zero in the $\tilde{M}=\tilde{M}_{monopole}$ limit
as with the Schwarzschild black hole. Consider the evolution of a RN black hole
with $\tilde{r}_{H}>\tilde{r}_{H,max}$ by evaporation process. 
In GR in this figure, it loses mass, and
when the solution approaches the minimum of the inverse temperature,
its heat capacity changes discontinuously. [Note that its specific heat
changes its sign continuously.] This is the second order phase transition.
The black hole loses its mass further and reaches the point $B$.
Below this mass, a RN black hole becomes unstable and the solution
traces the monopole black hole branch. The nontrivial structure
comes out of the event horizon gradually. This is also
the second phase transition. At this point, however, both 
the heat capacity and the specific heat change discontinuously.
After that, the evaporation
continues until the event horizon disappears. Finally, the 
regular monopole may remain as a remnant. 
For $\omega=0$, however, it experiences a different evolution, because before it 
reaches the minimum of the inverse temperature, it reaches the point $B$ 
and traces the monopole black hole branch continuously.

Furthermore, the $\tilde{\lambda}< \tilde{\lambda}_{crit}^{(ii)}$ case is  different. 
Fig.~11(b) shows the inverse temperature of the solutions with 
the $\tilde{\lambda}=0.1$,
$\tilde{v}=0.1$. In the $\omega=-1.4$, $-1.0$  and $0$ case,
the RN black hole evolves to the point B as in the above case.
However, since the RN black hole becomes unstable, the solution jumps
to the stable monopole branch, increasing its entropy and temperature
discontinuously (arrows in Fig.~11(b)). This is the first order phase 
transition. The nontrivial structure comes out suddenly (or 
catastrophecally)
from the horizon.
In the $\omega=\infty$ case, the solution experiences
the second order phase transition at $\tilde{M}=2\tilde{Q}/\sqrt{3}$
before the first order phase transition.  As shown 
above, the type and number of the phase transitions depend
on the physical parameters. 
This property is more important for the stability of the 
solution in
the system surrounded by a heat bath.
In that system, stability change occurs at the point where
the heat capacity diverges\cite{Katz}.
Another thing that we should note is that the monopole black hole
objects are forced to be produced by this process if the accretion to the
black hole is very little.

%-v_max
Until now, we have studied the solution in the rather small value of $\tilde{v}$
corresponding to the left side of Fig.~5. and found 
the region (ii) in BD theory. Next we consider the $\tilde{v}_{max}$
in BD theory.
By the same method in the gravitating monopole case, we found
$\tilde{v}_{max}$. Fig.~4 shows 
the existence region of the black hole solution ($\tilde{r}_{H}=0.4$) in 
$\tilde{v}$-$\tilde{\lambda}$ plane for $\omega=0$ and $\infty$. 
In the left region of each dotted line, there are black hole
solutions. When we take $\tilde{\lambda}$ as larger, 
the dependence of $\tilde{v}_{max}$ becomes smaller. 
We can not see the intrinsic difference also in this diagram. 
It is different to clarify the detailed structure
of region (i) in BD theory because of the fine tuning of the asymptotic
value of the BD field. In GR the region (i) disappears at 
$\tilde{\lambda} = \tilde{\lambda}^{(i)}_{crit} \sim 0.287$. As
discussed before, since $\tilde{\lambda}$ does not depend 
on $\phi_0$, $\tilde{\lambda}^{(i)}$ would not be changed
much even in BD theory. However, since the gravitational effect becomes 
strong around $\tilde{v}=\tilde{v}_{max}$, the BD scalar field may 
affect the value of $\tilde{\lambda}^{(i)}_{crit}$. It 
needs further analysis to say definite thing.

%%%%%%%%%%%%%%%%%%%%%%%%%%%%%%%%%%%%%%%%%%%%%%%%%%
%%%%%%%%%%
%                                                          %
\section{Conclusion}    %
%                                                          %
%%%%%%%%%%%%%%%%%%%%%%%%%%%%%%%%%%%%%%%%%%%%%%%%%%
%%%%%%%%%%

We found the self-gravitating monopole and its black hole solution of monopole in BD theory 
and compared them with those in GR.  
We clarified the following important aspects. 
The configuration of the YM potential and the Higgs fields of a self-gravitating 
monopole and a monopole black hole hardly depend on the BD parameter
in most of the mass range, which shows the structure resembles the 't Hooft-Polyakov
monopole.
We found the parameter region (ii) in Fig. 5, where two monopole black hole 
solutions exist for fixed horizon radius. 
Each solution branch has a different dependence for the BD parameter. 
For the higher entropy branch, the configurations of the fields 
hardly depend on the BD parameter, while
the configuration of the fields changes to those of RN black hole in the
$\tilde{v} \to \tilde{v}_{max}$ limit for the lower entropy branch.
However, we can not find the region (i) because fine tuning is needed
to satisfy the boundary condition of the BD field at infinity.
One of the characteristic properties in BD theory is lessening the effective
global gauge charge. By this effect, the $\tilde{M}$-$\tilde{r}_h$ diagram
(Fig.~7) shifts to the left side, i.e., the mass of the solution becomes
small when $\omega$ gets small. From this diagram, we can derive information
about stability by applying catastrophe theory. As in the GR case,
the high (low) entropy branch is stable (unstable). And we found that the BD parameter
is not an intrinsic parameter but a dummy parameter in catastrophe theory.
The effective global charge also 
affects the configuration for the 
high entropy branch, and does affect the existence region of the solutions in parameter regions. 
We investigated a boundary value of $\tilde{v}$ above which there exist no 
static solution for $\tilde{r}_{h}=0.4$ and a self-gravitating solution.
It is interesting that the allowed parameter region is extended in BD theory.
We also discuss the thermodynamical properties of the black hole solutions.
The number and the order of the phase transitions depend complicatedly 
on the theoretical parameters.

These aspects may give us some important information about monopole black holes in other 
theories of gravity. In particular, we can anticipate the results in 
dilaton gravity because of its similarity to BD theory. Though we can not 
see the intrinsic qualitative difference form GR, some effective theory of 
unified theory may cause 
important effects for monopole \cite{Harvey} or its black hole solution,
which may predict some information
to the remnant of the Hawking evaporation. 

There are some issues which we should comment on. In this paper,
we transformed the system from the original BD frame to the Einstein frame and
discussed only in the latter frame. However, the solutions in the
former frame are easily obtained from our solution by the
inverse transformation of Eq.~(\ref{B2}). For example,
$M$-$r_H$ diagram in the BD frame is obtained as follows. 
In the BD frame, a gravitational mass $M_{BD}$ is defined by use of the scalar 
mass as\cite{Will} 
%%%%%%%%%%%%%%%%
\begin{eqnarray}
M_{BD}=M+M_{s}
\ . \label{gravitational mass of BD frame}
\end{eqnarray}
%%%%%%%%%%%%%%%%
The $\tilde{M}_{BD}$-$\tilde{r}_{H}$ diagram is shown in Fig.~12,  
where we normalize as $\tilde{M}_{BD}=Gev M_{BD}$. Because of the existence 
of a scalar field, a cusp structure disappears at the points $A$. 
This may show that $\tilde{M}_{BD}$ is not appropriate to 
a control parameter of catastrophe theory\cite{Tamaki,Koga}. 
Since the solution in a high entropy branch has a 
larger scalar mass than that in a low entropy branch as shown in Fig.~10, 
the high entropy branch slides to the right of the lower branch in the BD frame.

We note the relation between our results and other topological defects such as a 
global monopole or a cosmic string in BD theory or in more 
general scalar-tensor theories. For global monopole, it is investigated that 
a massless dilaton field or the BD scalar field affects its structure, and is discussed 
that there is no solution except the special value of the coupling constant\cite{Barros}. 
Similar arguments are seen for a cosmic string in \cite{Gundlach} where 
they discussed that ``wire approximation" which can be valid in GR will not be used. 
For a massive dilaton field, though both global monopole and cosmic strings 
would not be affected seriously compared to the massless case, it is discussed that it enables us to decide the dilaton 
mass from observation\cite{Damour}. As for the texture, though exact results are 
almost unknown but Dando using self-similar ansatz showed that  
regular solution exists even in a massless dilaton 
case for a suitable choice of coupling constant\cite{Dando}. 
Contrary to these cases, both the self-gravitating monopole 
and its black hole solution exist for any value of the BD parameter in our model. 
This would be due to the asymptotic mild behavior of the BD scalar field. 

We finally note that much work on the monopole solution has been 
developing from the context of superstring theory\cite{Callan}.  
Recently monopole solutions in flat space-time,
self-gravitating monopole solutions and its black hole solutions were obtained
in Born-Infeld type action\cite{Grandi}.
However, little has been clarified about them. It should be interesting
to investigate their structure, especially the internal structure around the 
central singularity, because of the remarkable form of its action.

\section*{ACKOWLEDGEMENTS}
We would like to thank J. Koga and T. Tachizawa for useful discussions. 
(T. T.)$^{2}$ are  thankful for  financial support from the JSPS. This work
was supported partially by a Grant-in-Aid for  Scientific
Research Fund of the Ministry of Education, Science and Culture
(Specially Promoted Research No. 08102010 and No. 09410217), by a 
JSPS Grant-in-Aid
(No. 094162), and by the Waseda University Grant  for Special
Research Projects.
%%%%%%%%%%%%%%%%%%%%%%%%%%%%%%%%%%%%%%%
%%%%%%%%%%%%%%%%%%%%%%%%%%%%%%%%%%%%%%%

\newpage
\noindent
%%%%%%%%%%%%%%%%%%%%%%
%%%%%%%%%%%figures  %%
%%%%%%%%%%%%%%%%%%%%%%
\begin{figure}[htbp]
\begin{tabular}{ll}
\segmentfig{8cm}{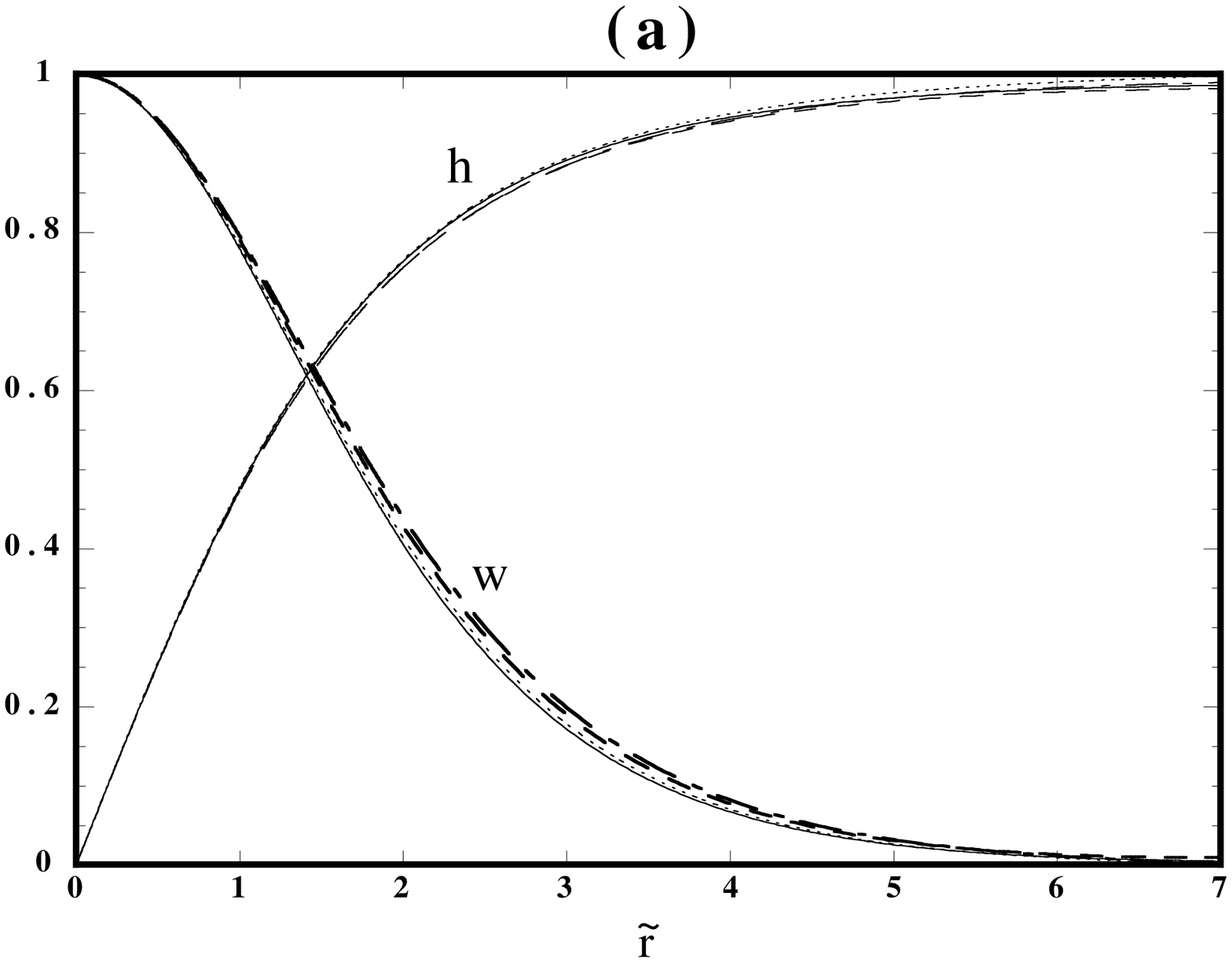}{}
\segmentfig{8cm}{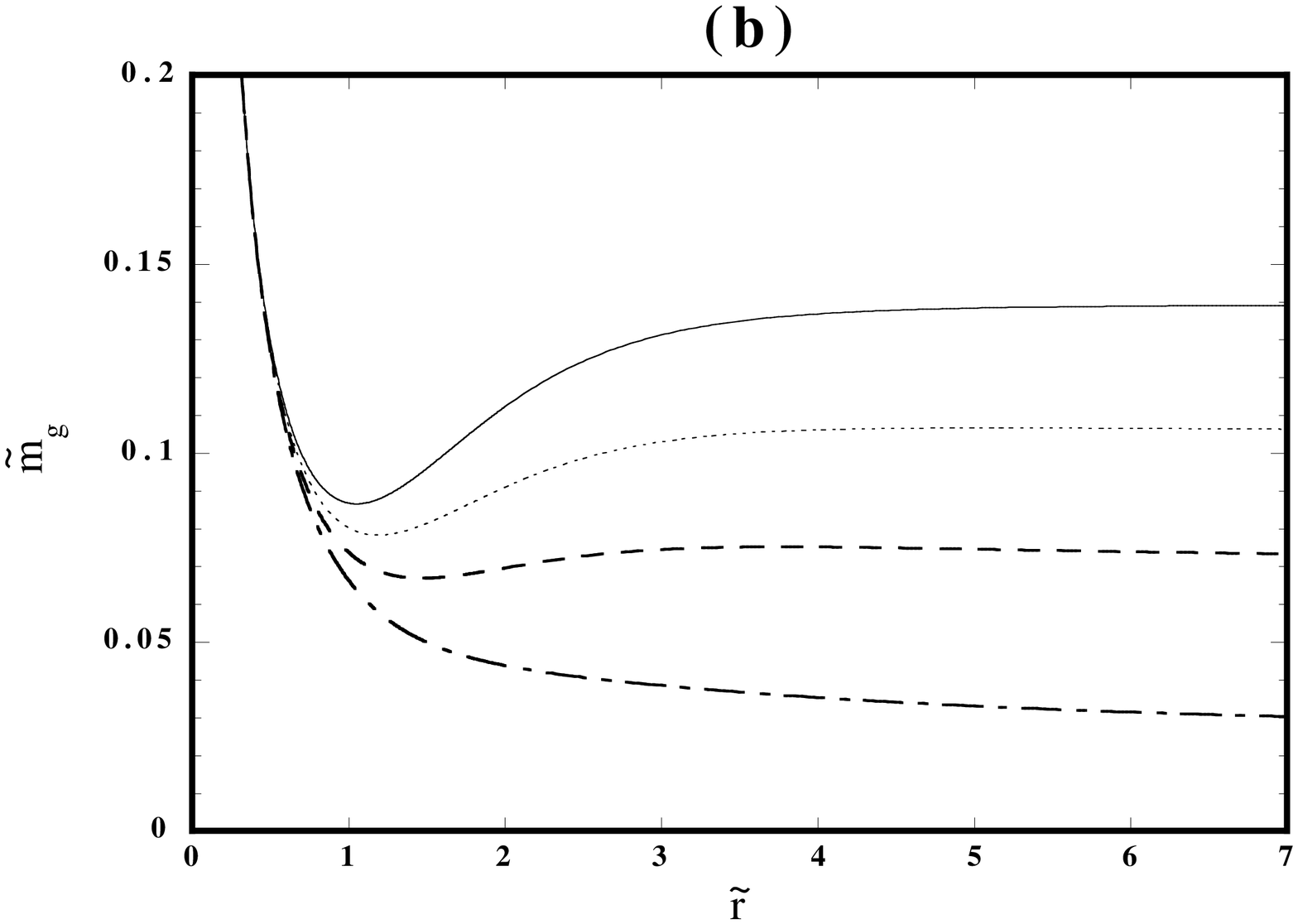}{}\\
\segmentfig{8cm}{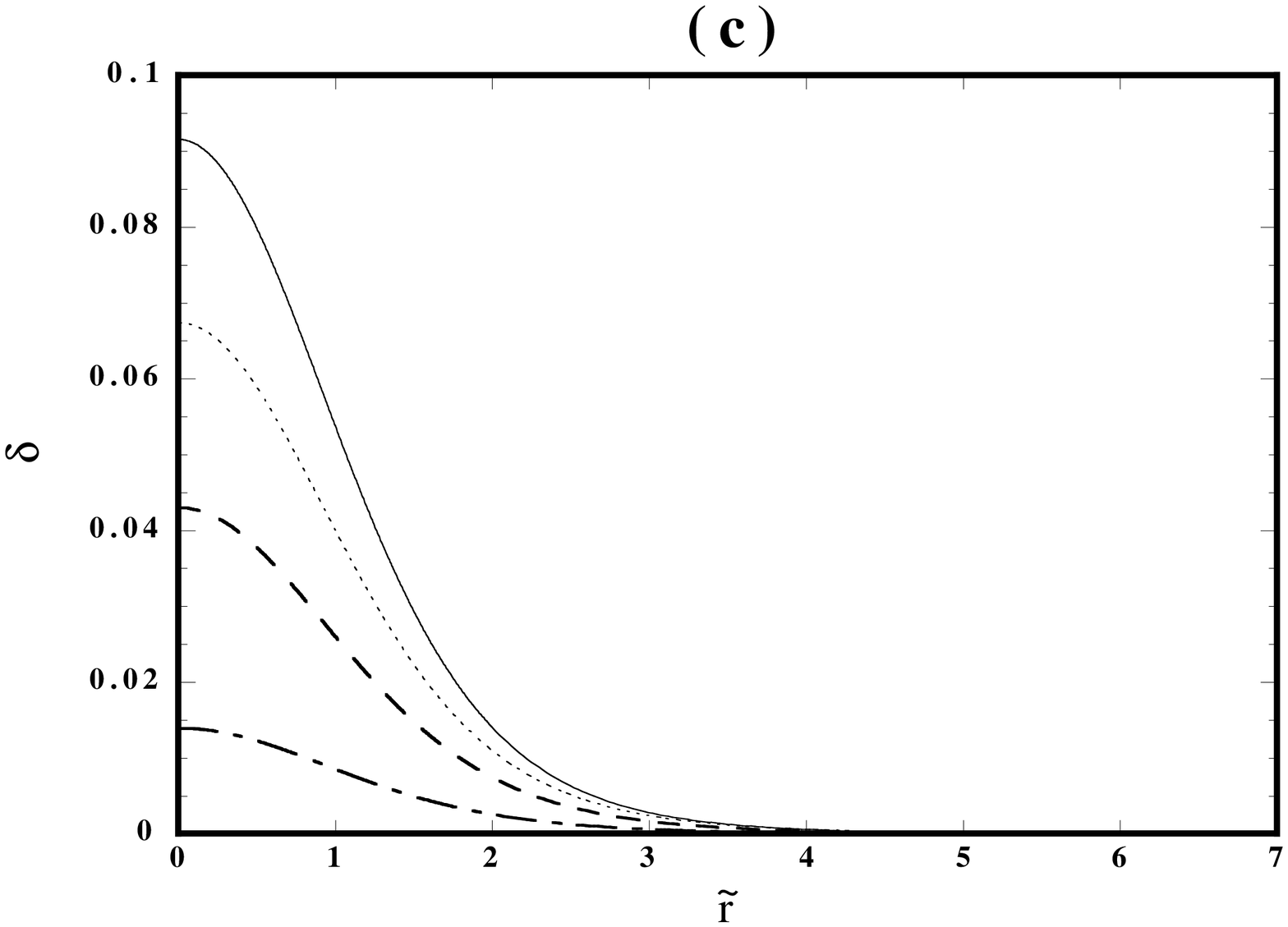}{}  
\segmentfig{8cm}{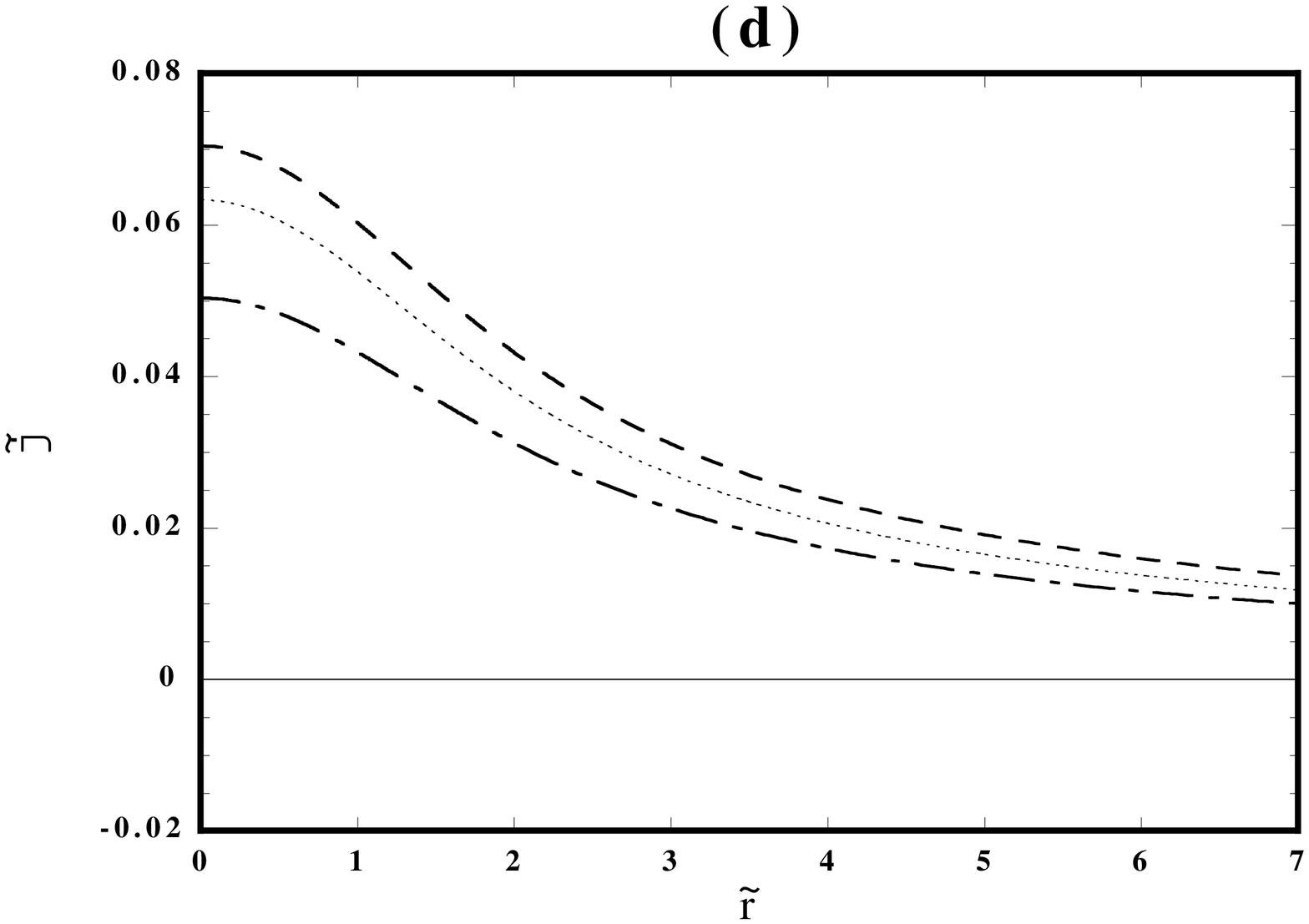}{}
\end{tabular}
\caption{ Field configurations of the self-gravitating monopole for 
$\tilde{v}=0.1$ and $\tilde{\lambda}=0.1$ ((a) $\tilde{r}$-$h$,$w$ 
(b) $\tilde{r}$-$\tilde{m}_{g}$ (c) $\tilde{r}$-$\delta$ (d) $\tilde{r}$-$\tilde{\varphi}$). 
We compared these distributions for $\omega =-1.4$, $-1$, $0$, $\infty$. 
Though metric functions and the scalar field depend on the BD parameter 
mainly due to the difference of the effective global 
charge, we can find that YM field and the Higgs field hardly do.   
\label{r-field-various-omega}   }
\end{figure}
%%%%%%%%%%%%%%%%%%%%%%%%
%%%%%%%%%%%%%%%%%%%%
\begin{figure}
\begin{center}
\singlefig{12cm}{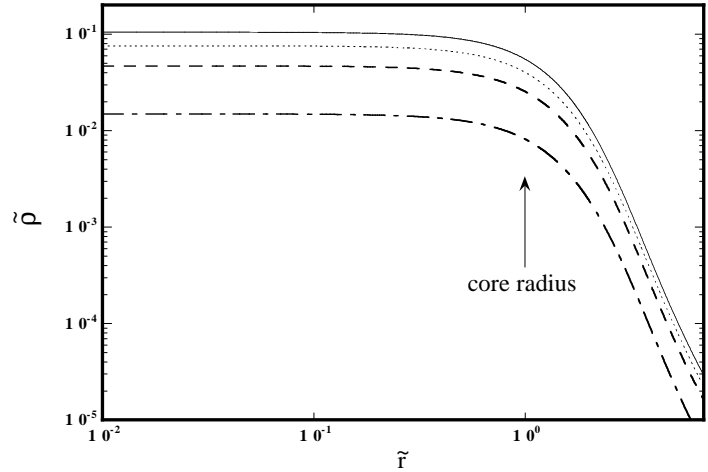}
\caption{The density distribution of the monopole solution for the same parameters as  Fig. 1. 
An arrow shows the core radius of the monopole which hardly depends on the 
BD parameter. The energy density decays as $1/\tilde{r}^{4}$ outside the core.  
\label{r-rho-various-omega} }
\end{center}
\end{figure}
%%%%%%%%%%%%%%%%%%%%
%%%%%%%%%%%%%%%%%%%%
\begin{figure}[htbp]
\begin{tabular}{ll}
\segmentfig{8cm}{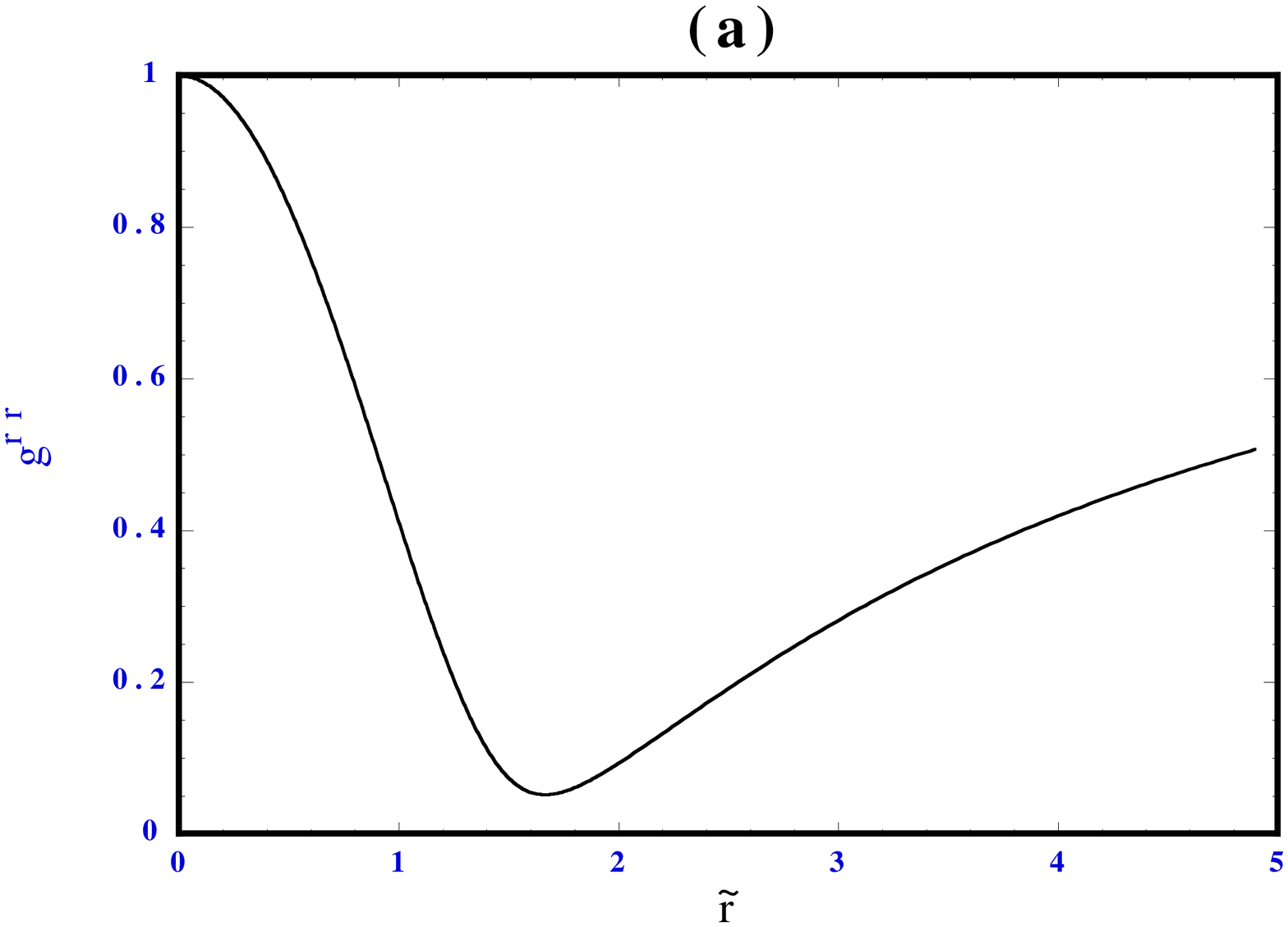}{}
\segmentfig{8cm}{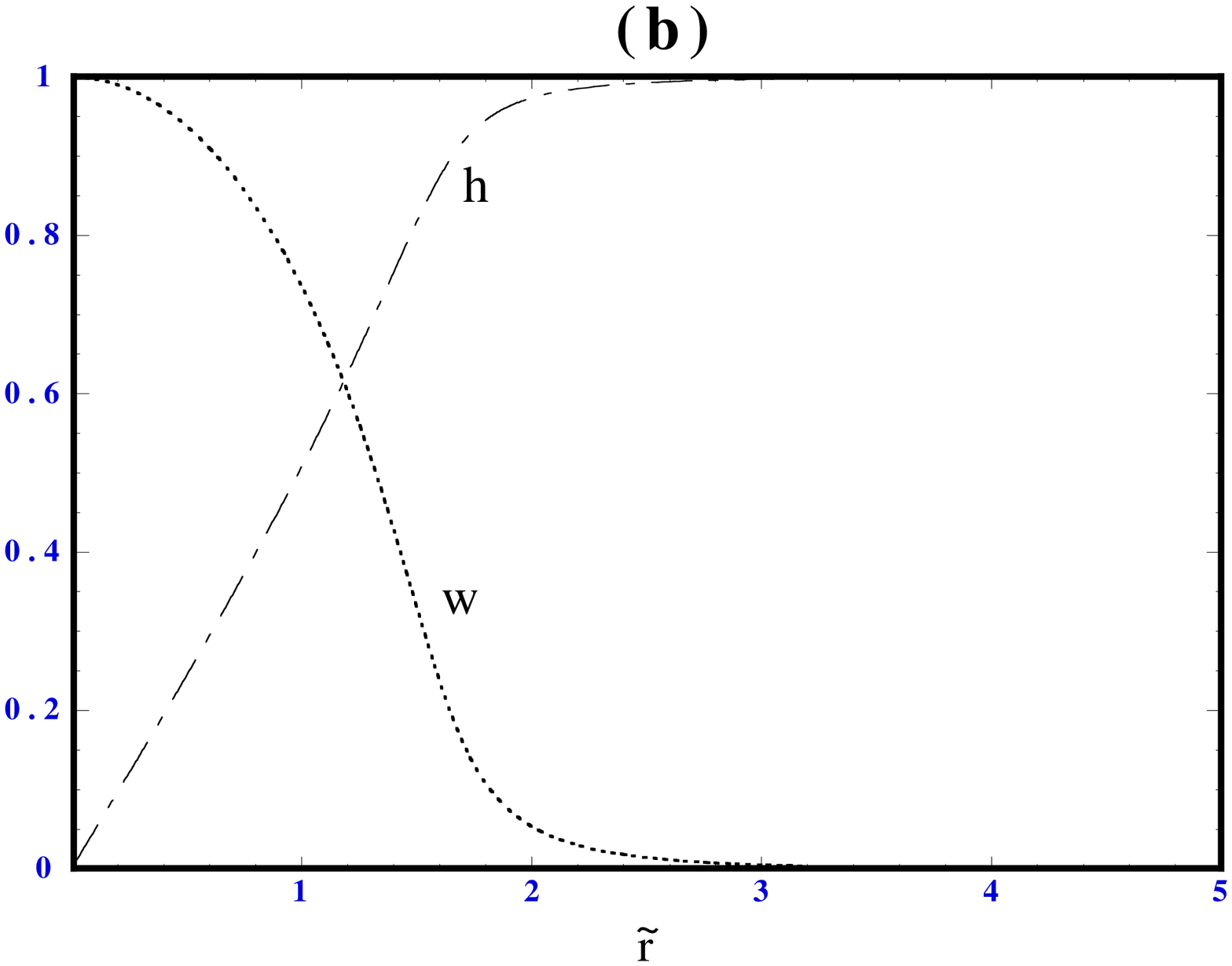}{}
\end{tabular}
\caption{ Field configurations of the self-gravitating monopole with $\tilde{\lambda}=0.2$ and $\omega =0$ 
near the $\tilde{v}_{max}$ ((a) $\tilde{r}$-$g^{rr}$ (b) $\tilde{r}$-$h$, $w$). 
Outside the minimum of $g^{rr}$, the solution can be approximated by RN black hole.   
\label{field-for-b0-around-vcrit}  }
\end{figure}
%%%%%%%%%%%%%%%%%%%%
\begin{figure}
\begin{center}
\singlefig{12cm}{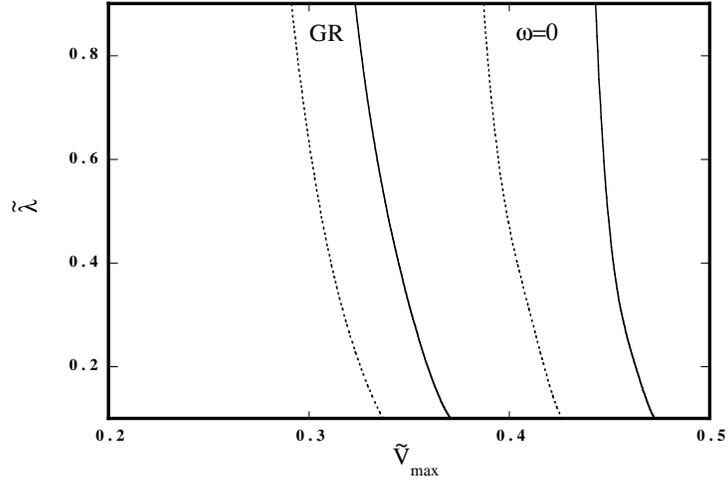}
\caption{The allowed region of the static solution for the self-gravitating monopole (the solid lines) 
and its black hole with $\tilde{r}_{H}=0.4$ (the dotted lines).
We compared these in the $\omega =0$ and $\omega =\infty$ cases. 
Static solutions can exist to the left side of each line.
For both solutions, $\tilde{v}_{max}$ becomes large in BD theory. 
For the self-gravitating monopole, the nontrivial structure is larger than  black hole's, so the change of 
$\tilde{v}_{max}$ from GR becomes larger than its black hole's.  
\label{Vmax-lam} }
\end{center}
\end{figure}
%%%%%%%%%%%%%%%%%%%%
%%%%%%%%%%%%%%%%%%%%
\begin{figure}
\begin{center}
\singlefig{12cm}{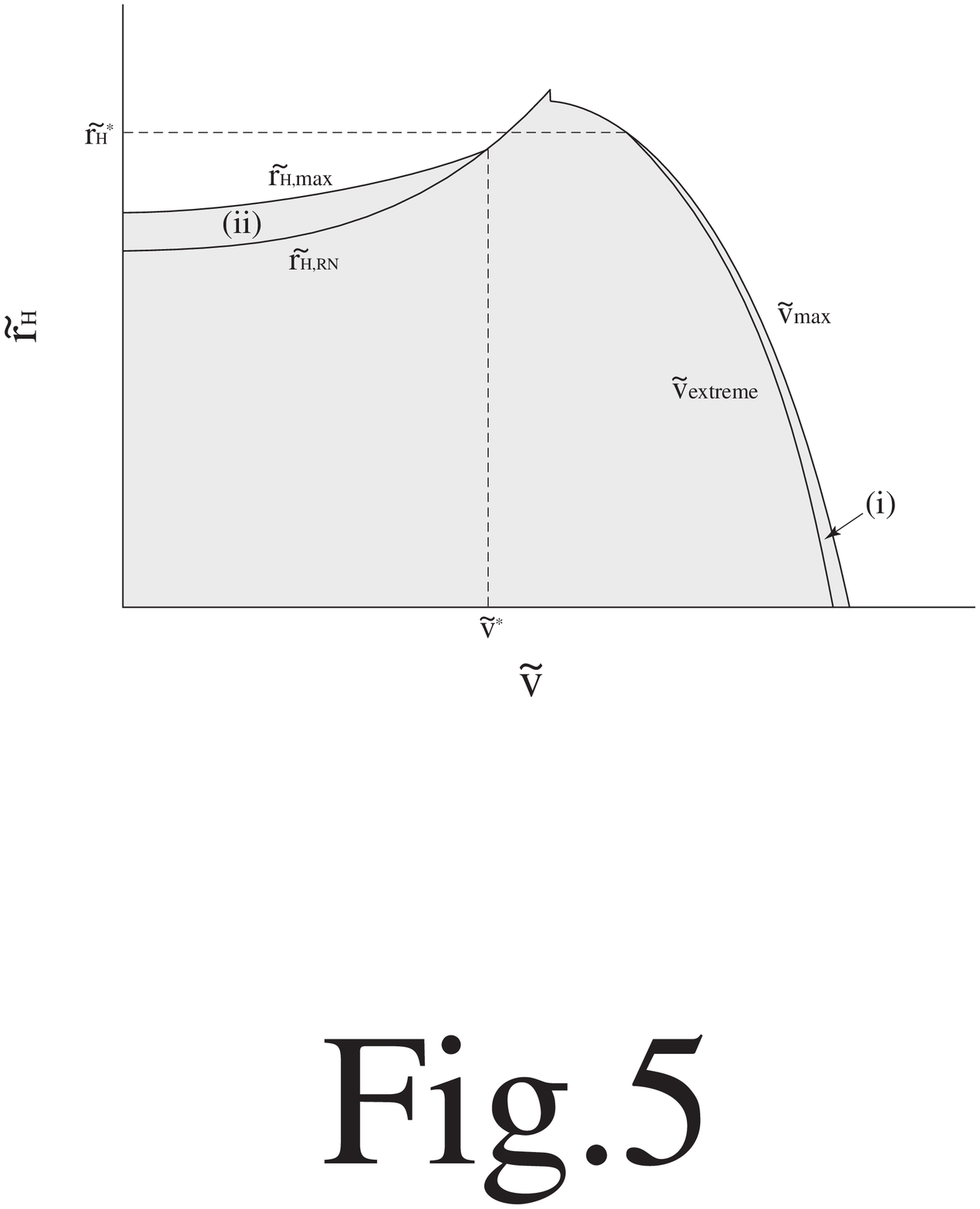}
\caption{Schematic diagram of the existence region in $\tilde{v}$-$\tilde{r}_{H}$ diagram. 
In the shadowed region, the nontrivial solution exists. For a fixed value of $\tilde{r}_{H}
<\tilde{r}_{H}^{\ast}$, there exists the region (i) ($\tilde{v}_{extreme}
<\tilde{v}<\tilde{v}_{max}$) where two nontrivial solutions with different mass exist. 
On the other hand, for a fixed value of $\tilde{v}<\tilde{v}^{\ast}$, 
there exists the region (ii) ($\tilde{r}_{H,RN}
<\tilde{r}<\tilde{r}_{H,max}$) where two nontrivial solutions with different mass exist. 
The region (i) and (ii) disappears for $\tilde{\lambda}>0.287$ and $\tilde{\lambda}>0.696$, 
respectively. 
\label{exregion-in-v-rh} }
\end{center}
\end{figure}
%%%%%%%%%%%%%%%%%%%%
%%%%%%%%%%%%%%%%%%%%%%
\begin{figure}[htbp]
\begin{tabular}{ll}
\segmentfig{8cm}{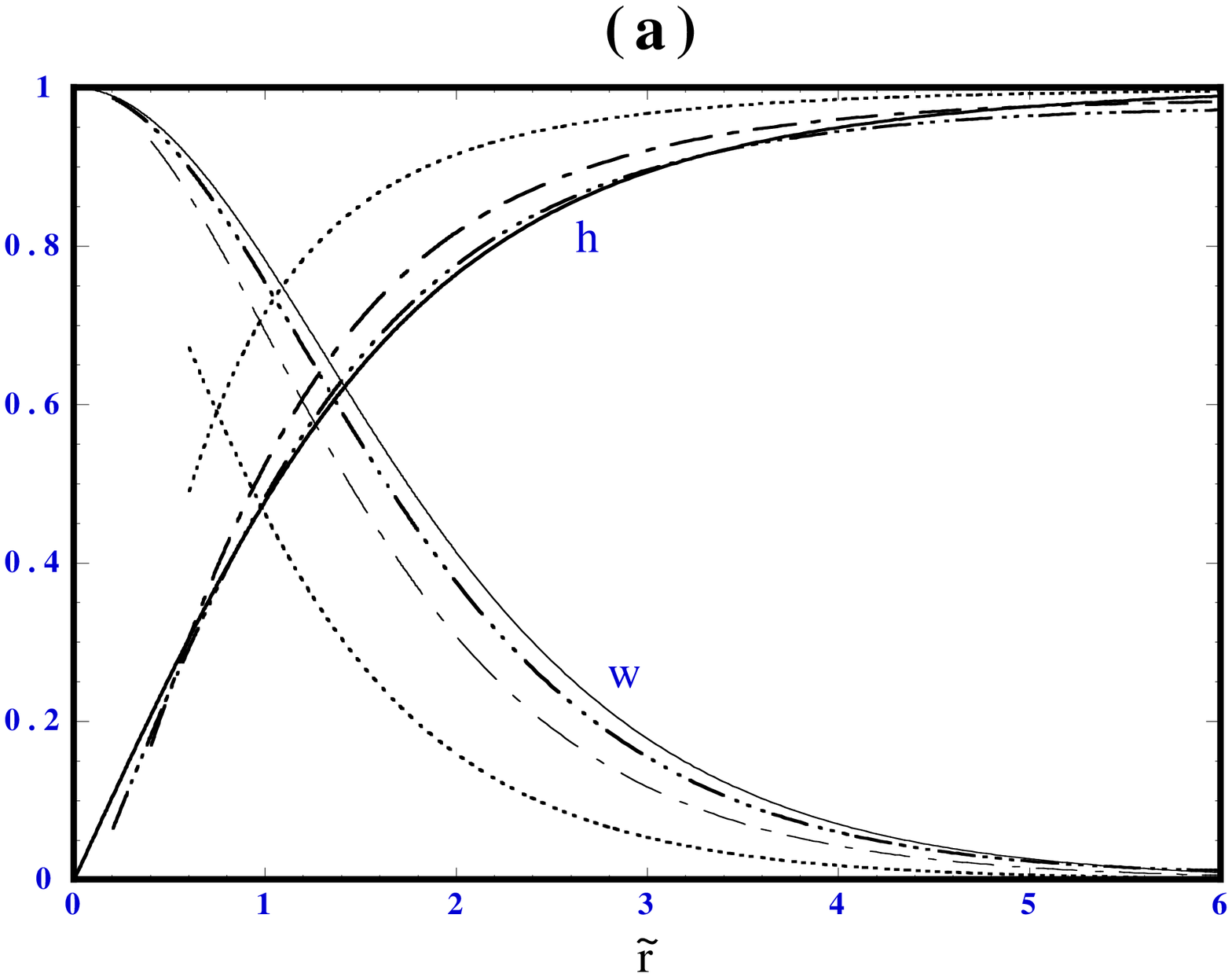}{}
\segmentfig{8cm}{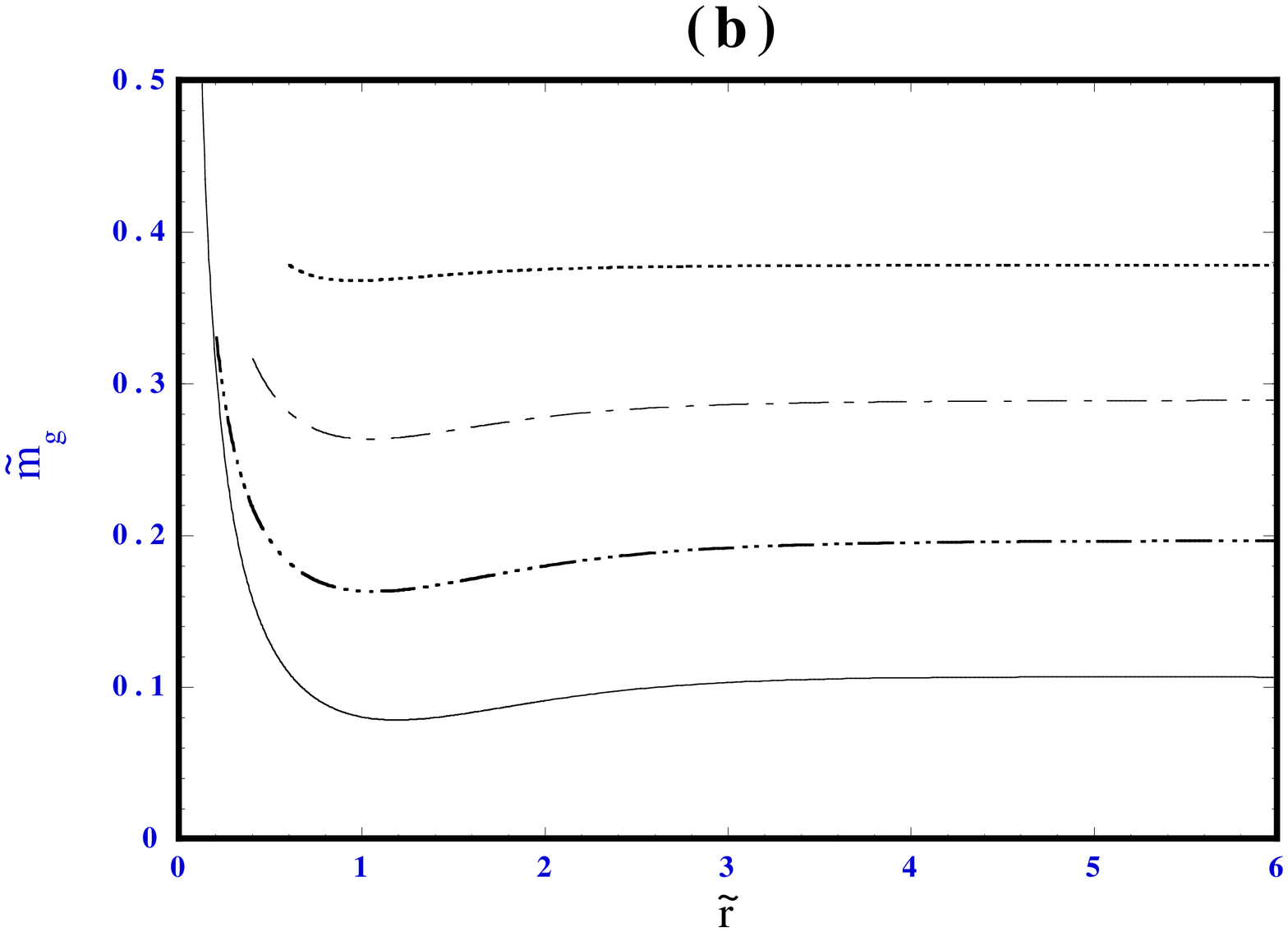}{}\\
\segmentfig{8cm}{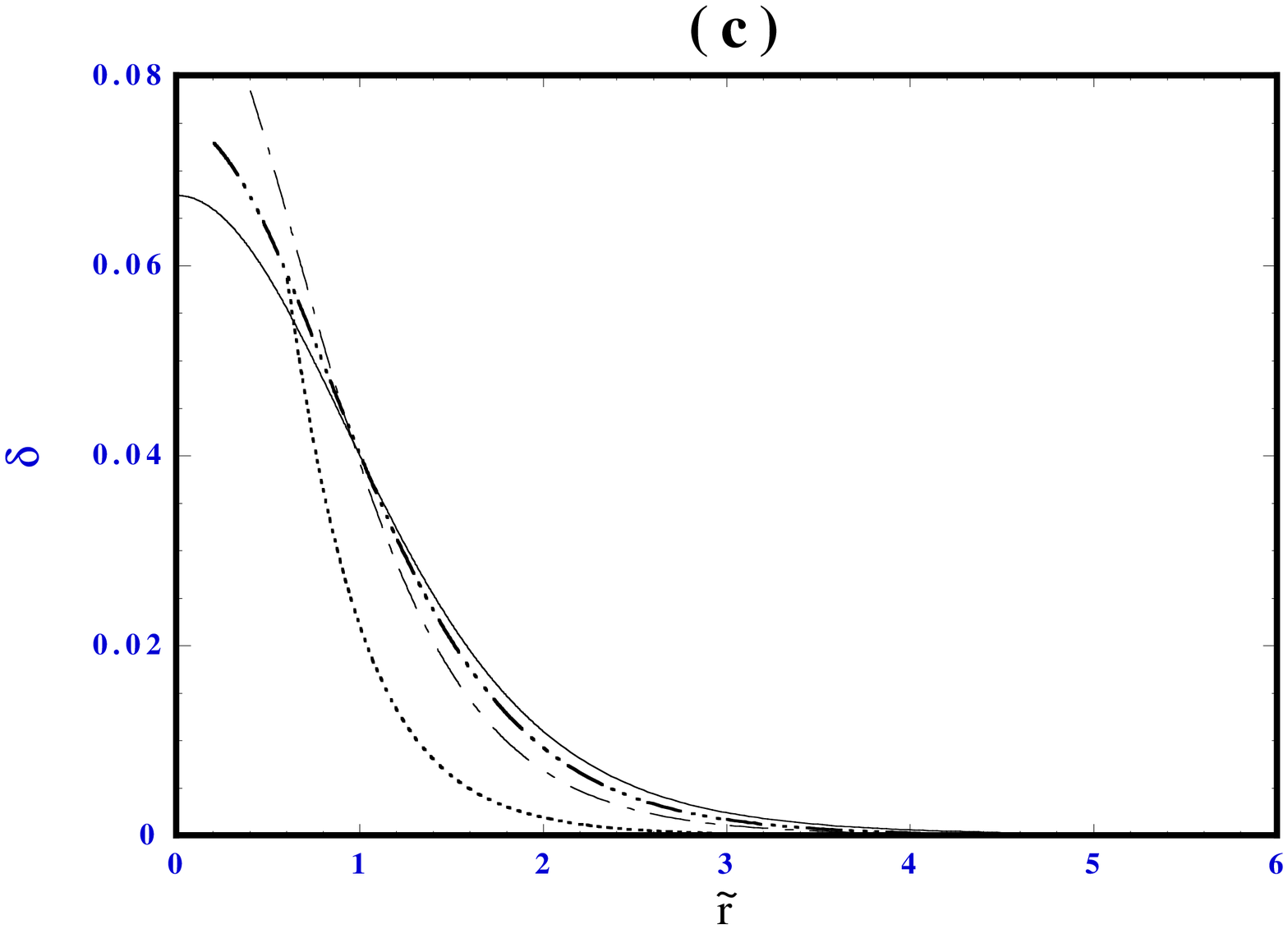}{}  
\segmentfig{8cm}{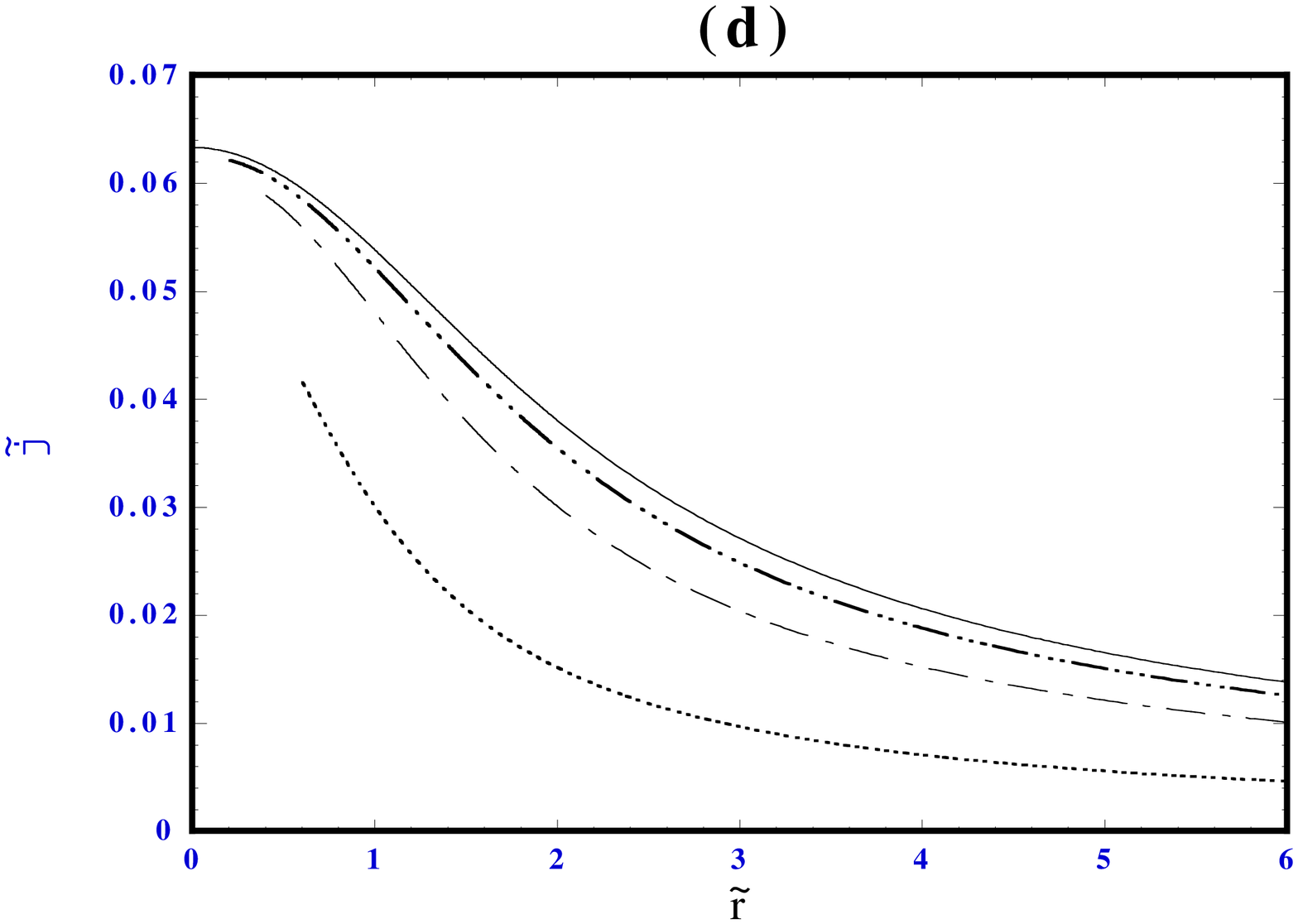}{}
\end{tabular}
\caption{  Field configurations of the monopole black hole for $\tilde{\lambda}=0.1$, 
$\tilde{v}=0.1$, $\omega =0$ and the horizon radius $\tilde{r}_{H}=0.2$, $0.4$, $0.6$
 ((a)$\tilde{r}$-$h$, $w$ (b) $\tilde{r}$-$\tilde{m}_{g}$ 
(c) $\tilde{r}$-$\delta$ (d)  $\tilde{r}$-$\tilde{\varphi}$). 
We also plot the self-gravitating monopole in the solid line (i.e., $\tilde{r}_{H}=0$)
for comparison. 
As the horizon radius becomes large, the nontrivial structure becomes small. 
So the variation of the BD scalar field becomes small. 
\label{fieldb0r02-06}  }
\end{figure}
%%%%%%%%%%%%%%%%%%%%%%%%
%%%%%%%%%%%%%%%%%%%%%%
\begin{figure}[htbp]
\begin{tabular}{ll}
\segmentfig{8cm}{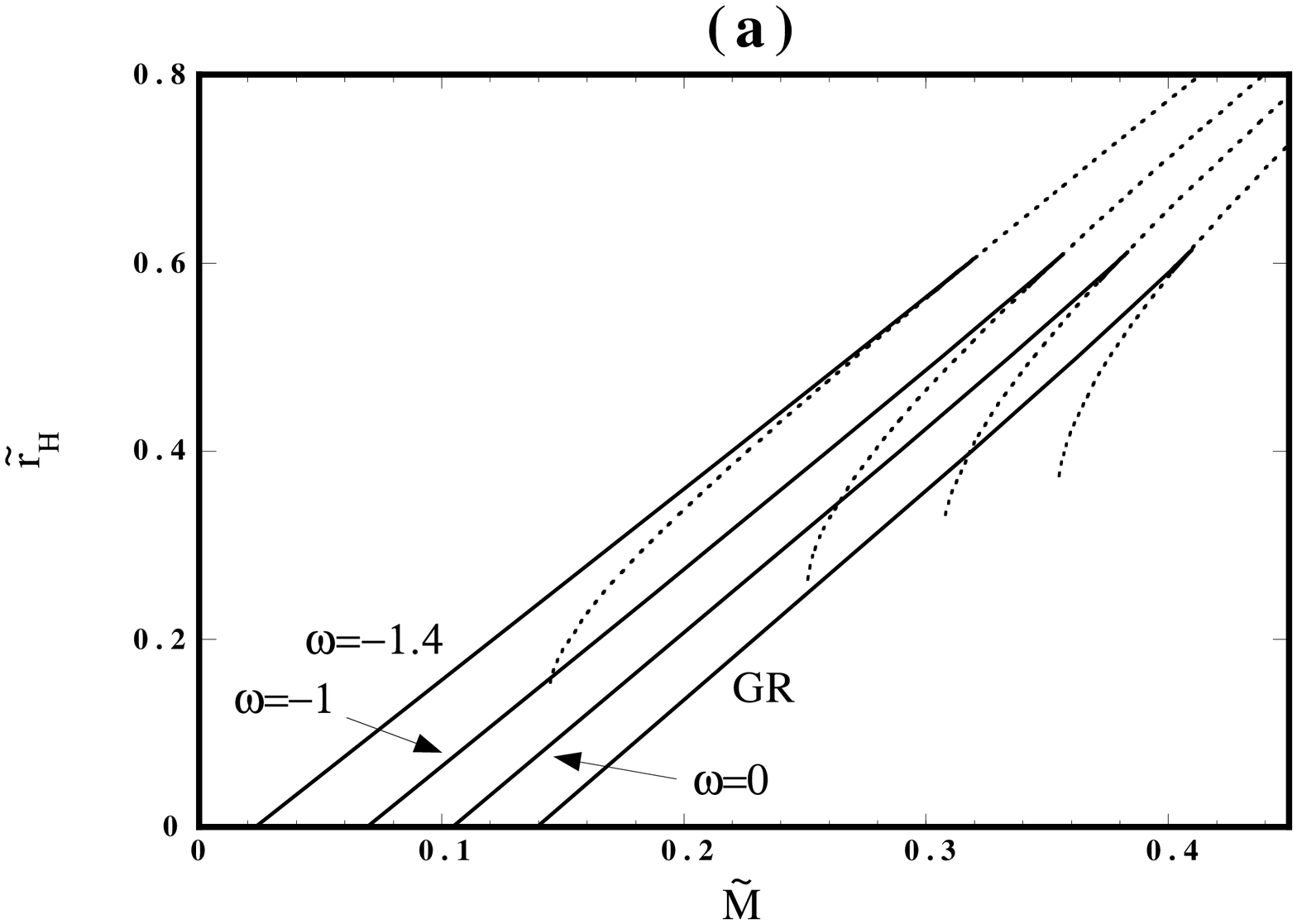}{}
\segmentfig{8cm}{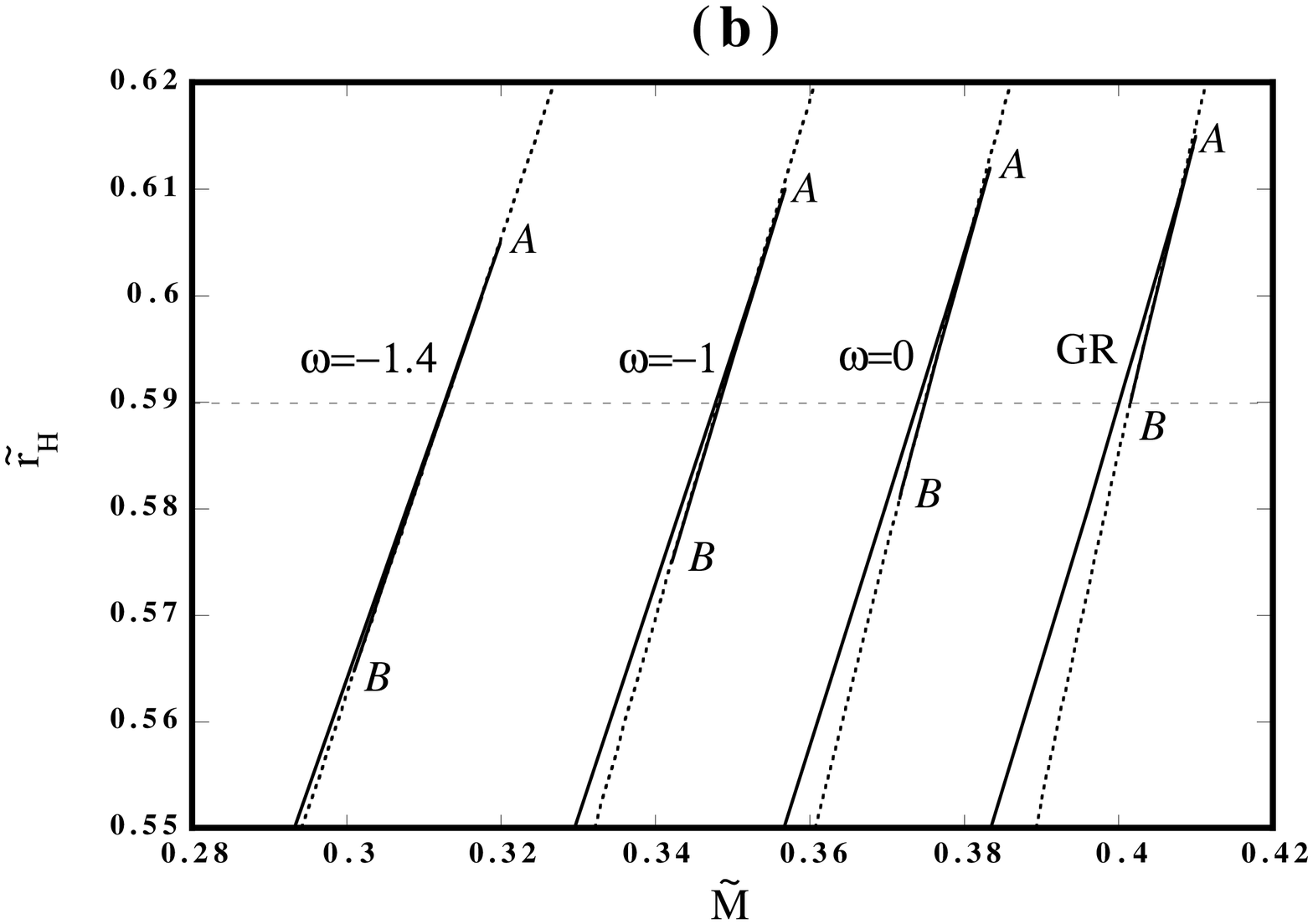}{}
\end{tabular}
\caption{  Mass-horizon radius diagram of the monopole black hole 
for $\tilde{v}=0.1$, $\tilde{\lambda}=0.1$ and $\omega =-1.4$, $-1$, $0$, $\infty$. 
((b) is a magnification of (a) around the maximum horizon radius.) 
The solid lines denote monopole black holes and the dotted lines denote RN black holes. 
Unlike the RN black hole, the monopole black hole does not have an extreme limit but 
has the $\tilde{r}_{H}=0$ limit, which corresponds to the self-gravitating monopole 
solution. The high entropy branch ends up with the maximum horizon radius and 
the maximum mass at point $A$, where a cusp structure appears.
The low entropy branch is connected to the RN black hole branch.
\label{M-rhGR&BDv0.1l0.1}    }
\end{figure}
%%%%%%%%%%%%%%%%%%%%%%%%
%%%%%%%%%%%%%%%%%%%%%%
\begin{figure}[htbp]
\begin{tabular}{l}
\segmentfig{10cm}{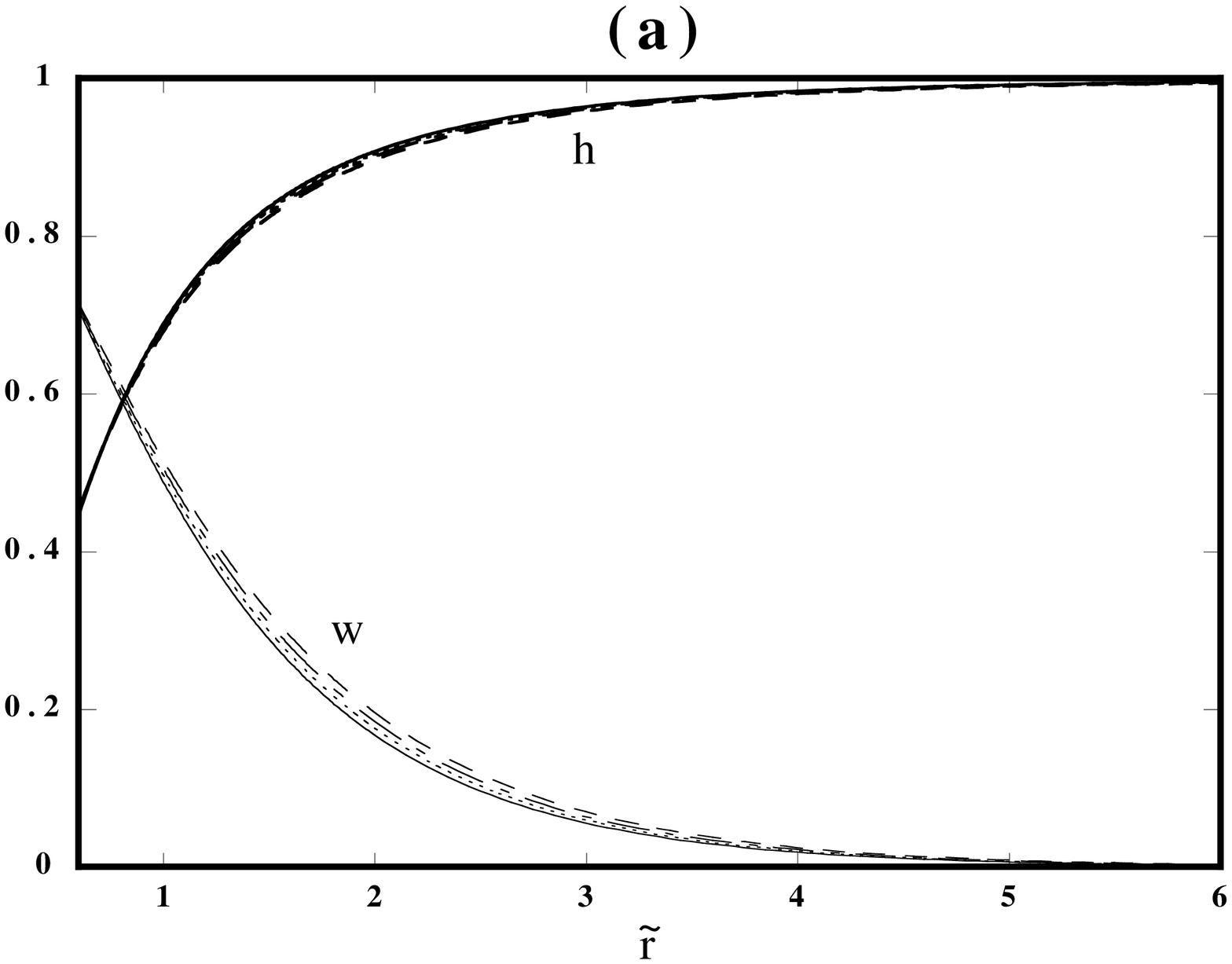}{}\\  
\segmentfig{10cm}{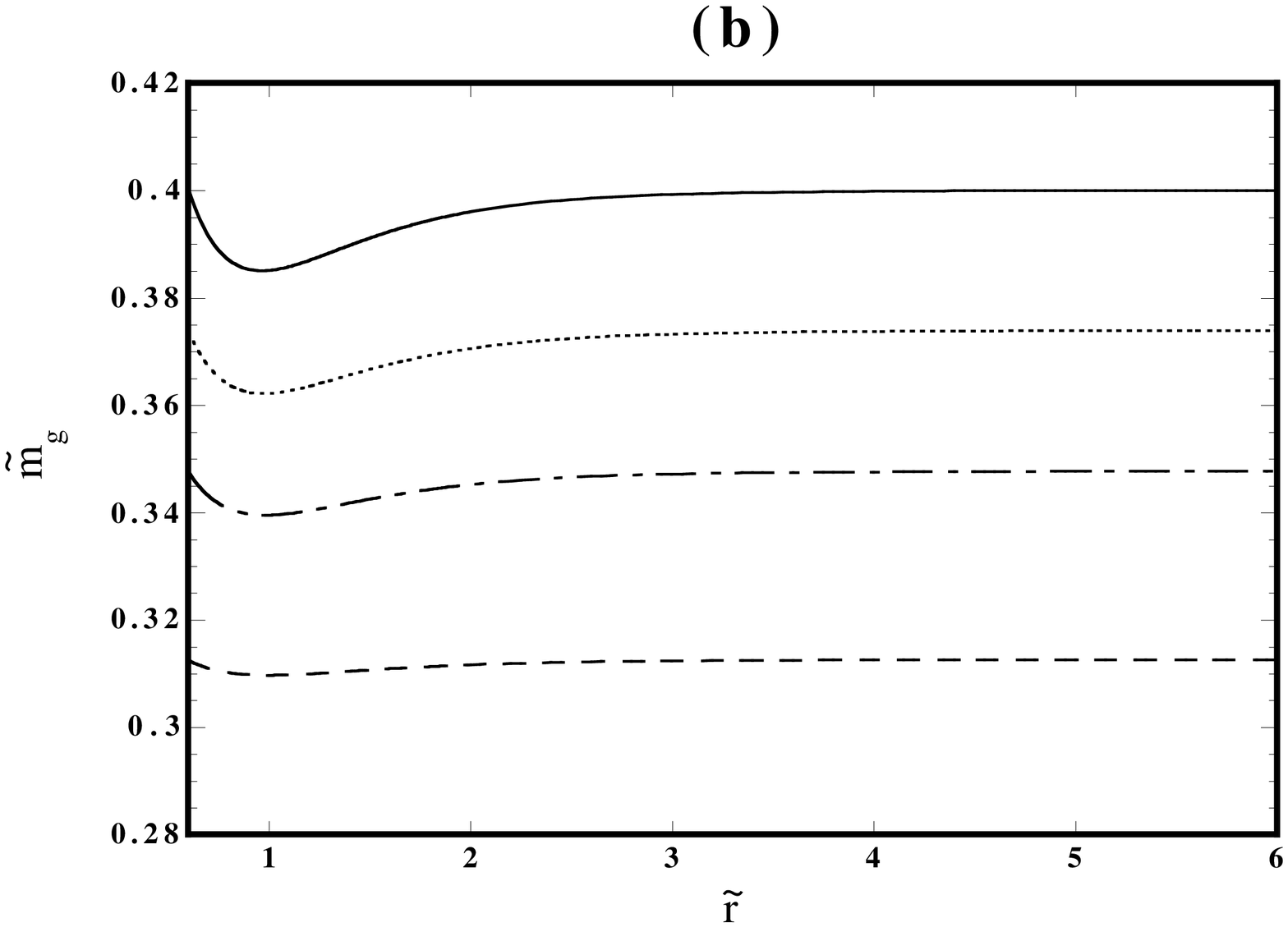}{}\\
\segmentfig{10cm}{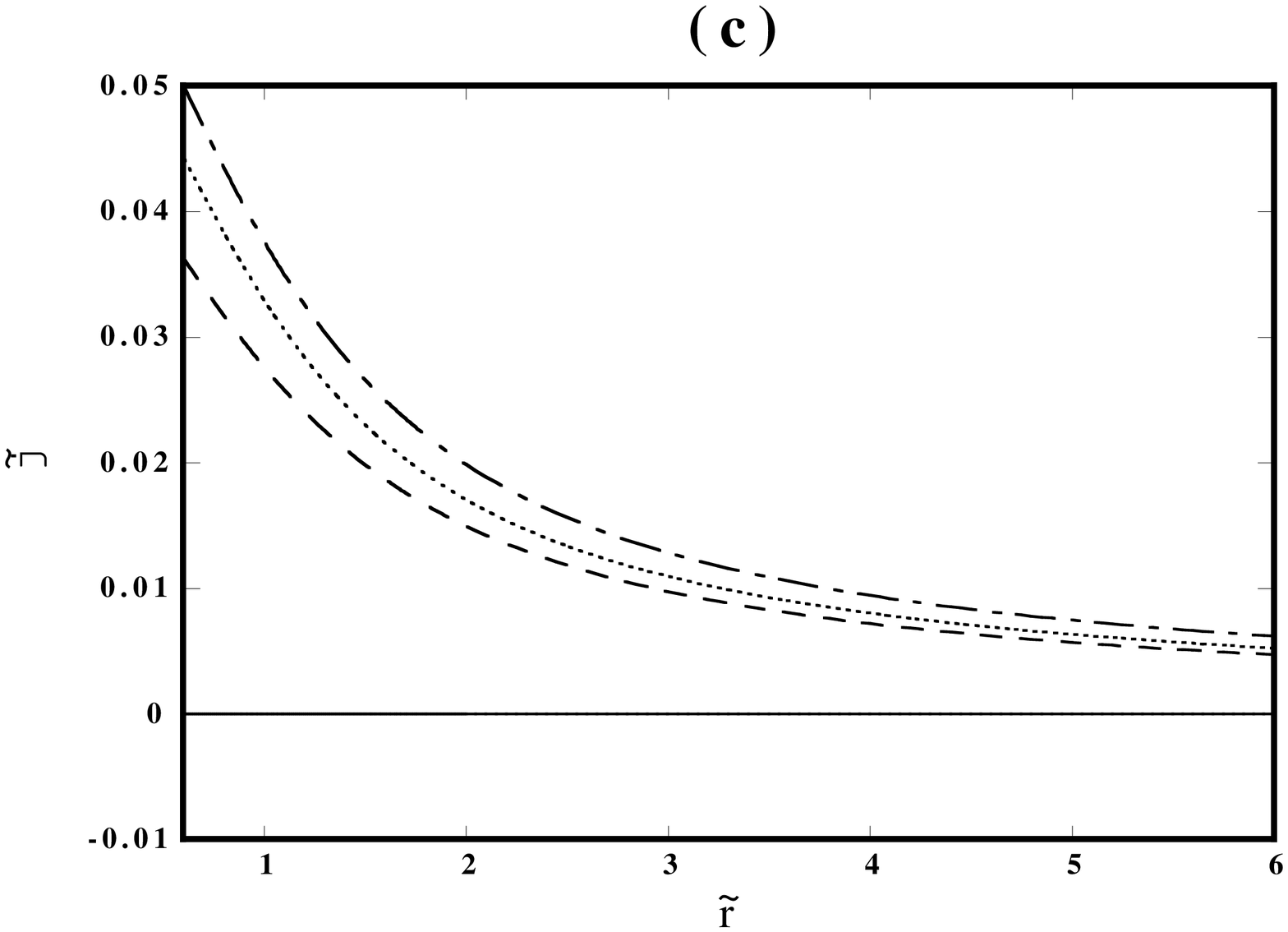}{}
\end{tabular}
\caption{  Field configurations of the solution in the high entropy branch for $\tilde{r}_{H}=0.59$, 
$\tilde{\lambda}=0.1$, $\tilde{v}=0.1$ and $\omega =-1.4$, $-1$, $0$, $\infty$
((a) $\tilde{r}$-$h$, $w$ (b)  $\tilde{r}$-$\tilde{m}_{g}$
(c) $\tilde{r}$-$\tilde{\varphi}$ ). The YM field and the Higgs field hardly depend on $\omega$.  
      \label{fieldr0.59st}    }
\end{figure}
%%%%%%%%%%%%%%%%%%%%%%%%
%%%%%%%%%%%%%%%%%%%%%%
\begin{figure}[htbp]
\begin{tabular}{l}
\segmentfig{10cm}{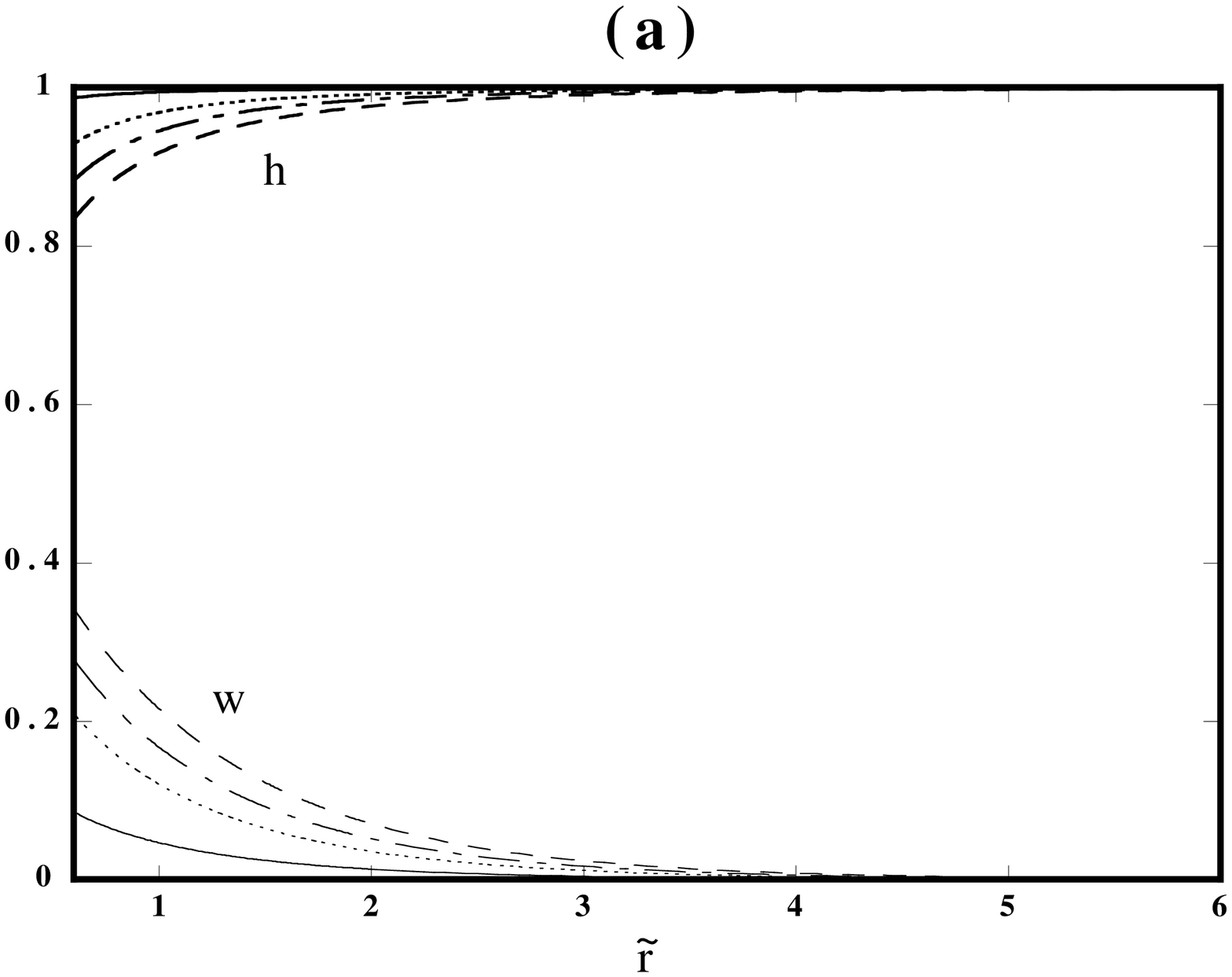}{}\\
\segmentfig{10cm}{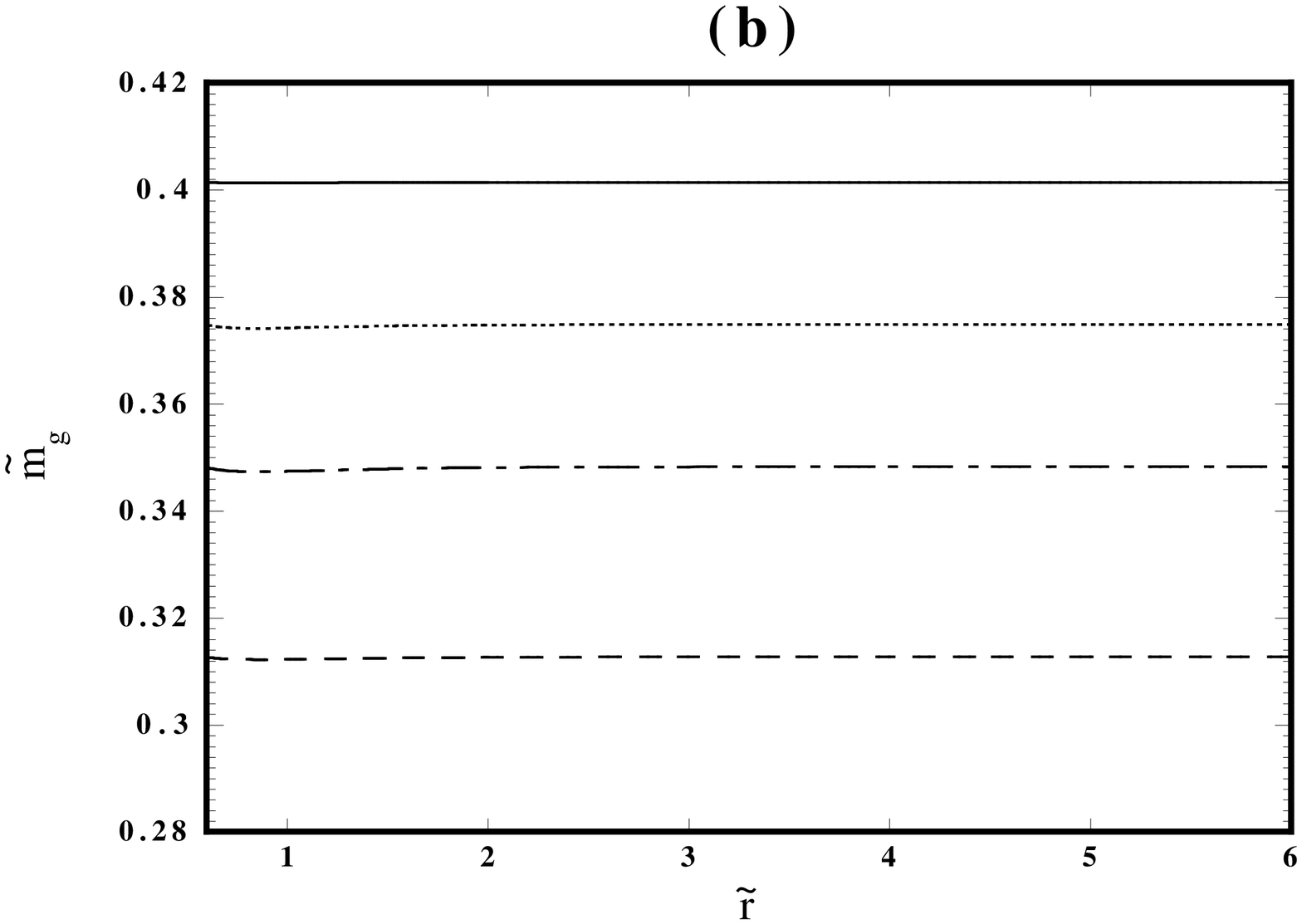}{}\\
\segmentfig{10cm}{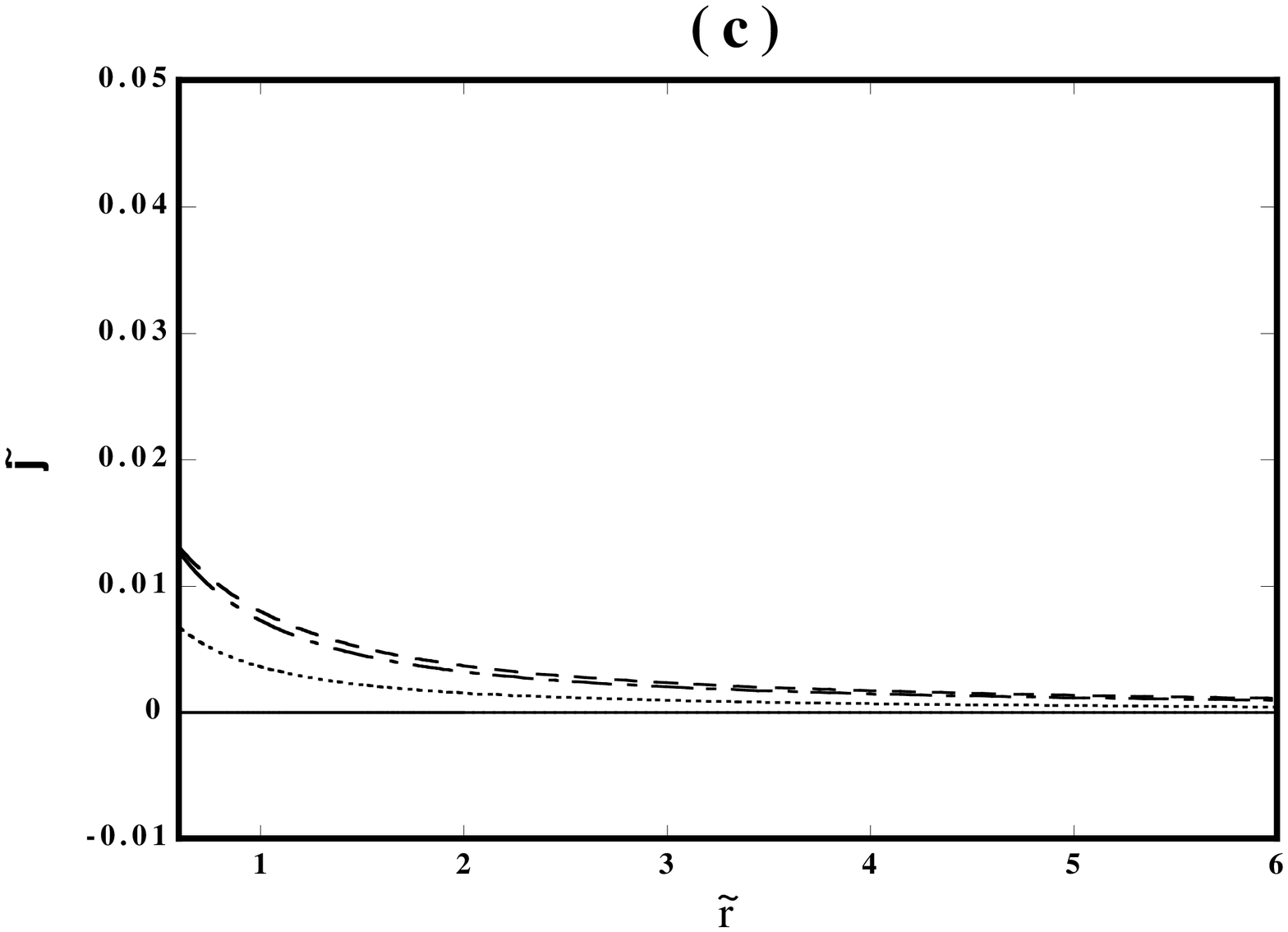}{}
\end{tabular}
\caption{  Field configurations of the solution in the low entropy branch for $\tilde{r}_{H}=0.59$, 
$\tilde{\lambda}=0.1$, $\tilde{v}=0.1$ and $\omega =-1.4$, $-1$, $0$, $\infty$
((a) $\tilde{r}$-$h$, $w$ (b) $\tilde{r}$-$\tilde{m}_{g}$
(c) $\tilde{r}$-$\tilde{\varphi}$ ). The YM field and the Higgs field depend on $\omega$ 
contrary to the high entropy branch.  It is because $\tilde{r}_{H}=0.59$ line crosses near the 
point $A$ for small $\omega$, but it crosses near the point $B$ for large $\omega$. 
      \label{fieldr0.59unst}     }
\end{figure}
%%%%%%%%%%%%%%%%%%%%%%%%
%%%%%%%%%%%%%%%%%%%%%%
\begin{figure}[htbp]
\begin{tabular}{ll}
\segmentfig{8cm}{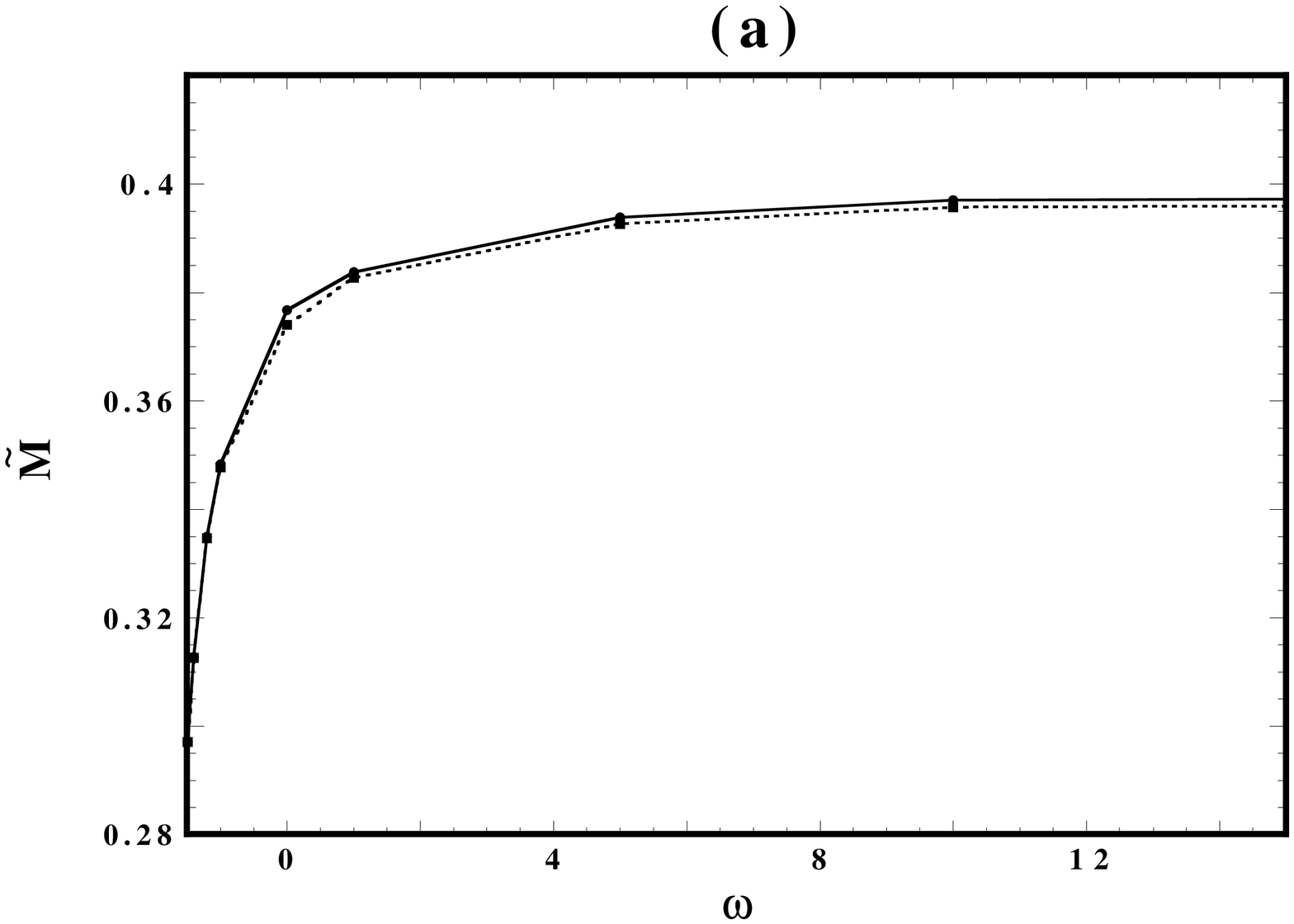}{}
\segmentfig{8cm}{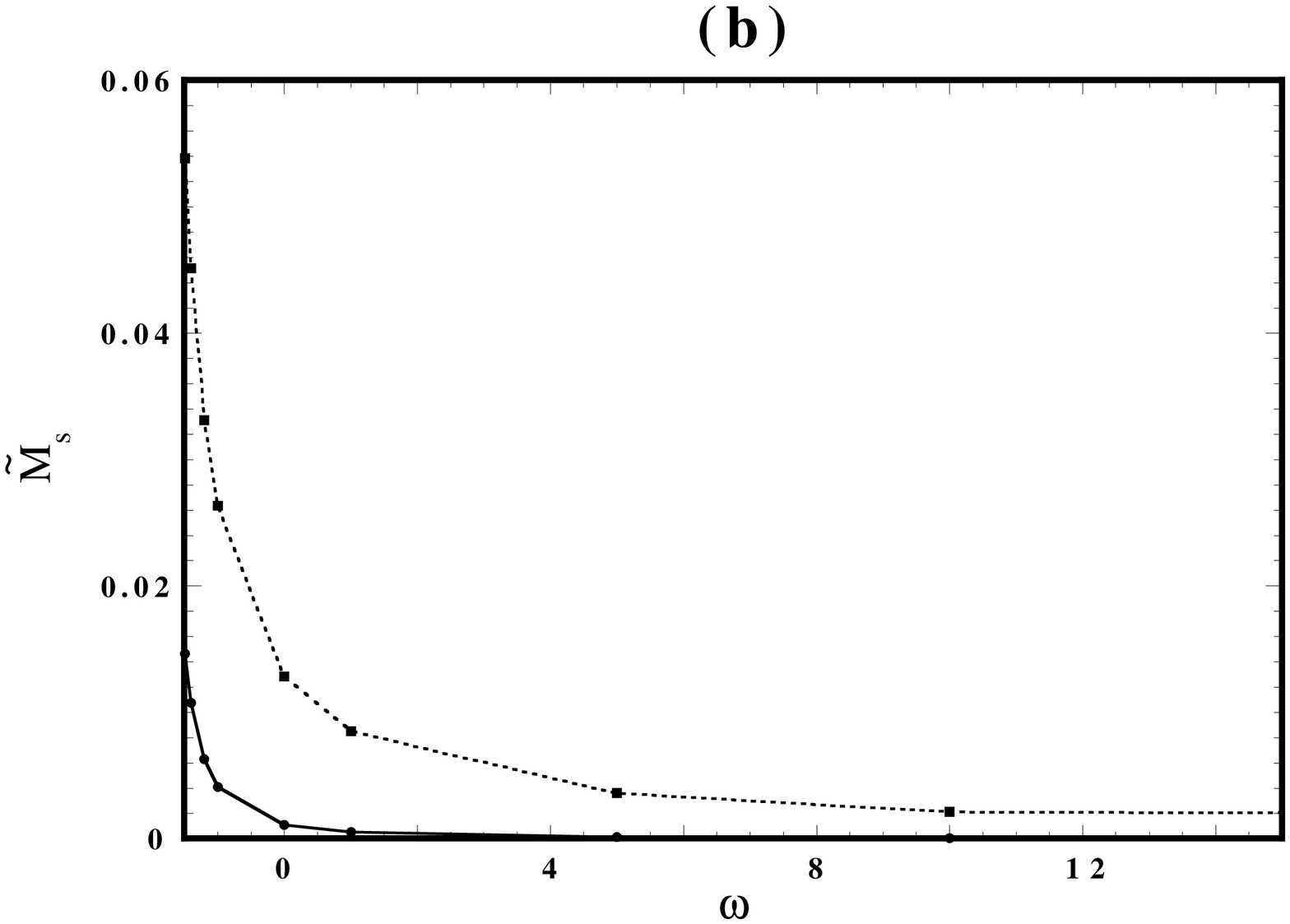}{}
\end{tabular}
\caption{  $\omega$ dependence of (a) the gravitational mass and (b) the scalar mass 
of  the monopole black hole for
$\tilde{v}=0.1$ and $\tilde{\lambda}=0.1$. 
We choose $\tilde{r}_{H}=0.59$ for which 
there exist two branches of monopole black holes for fixed $\omega$. 
Though dependence of $\tilde{M}$ in both branches can be approximated 
by a RN black hole, 
the effect of the nontrivial structure appears clearly for 
$\tilde{M}_{s}$. 
\label{omega-mass}     }
\end{figure}
%%%%%%%%%%%%%%%%%%%%%%%%
%%%%%%%%%%%%%%%%%%%%%%
\begin{figure}[htbp]
\begin{tabular}{ll}
\segmentfig{8cm}{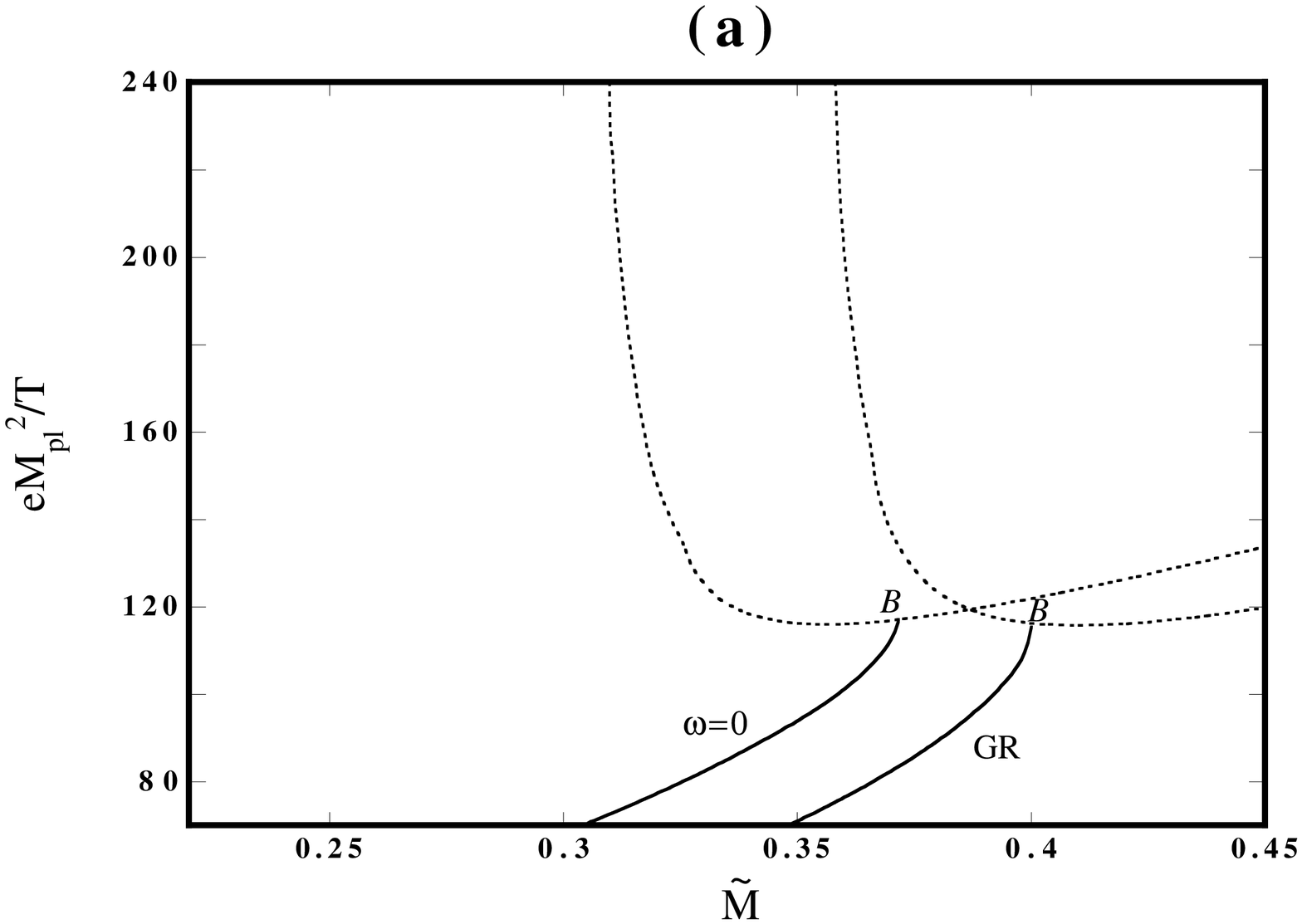}{}
\segmentfig{8cm}{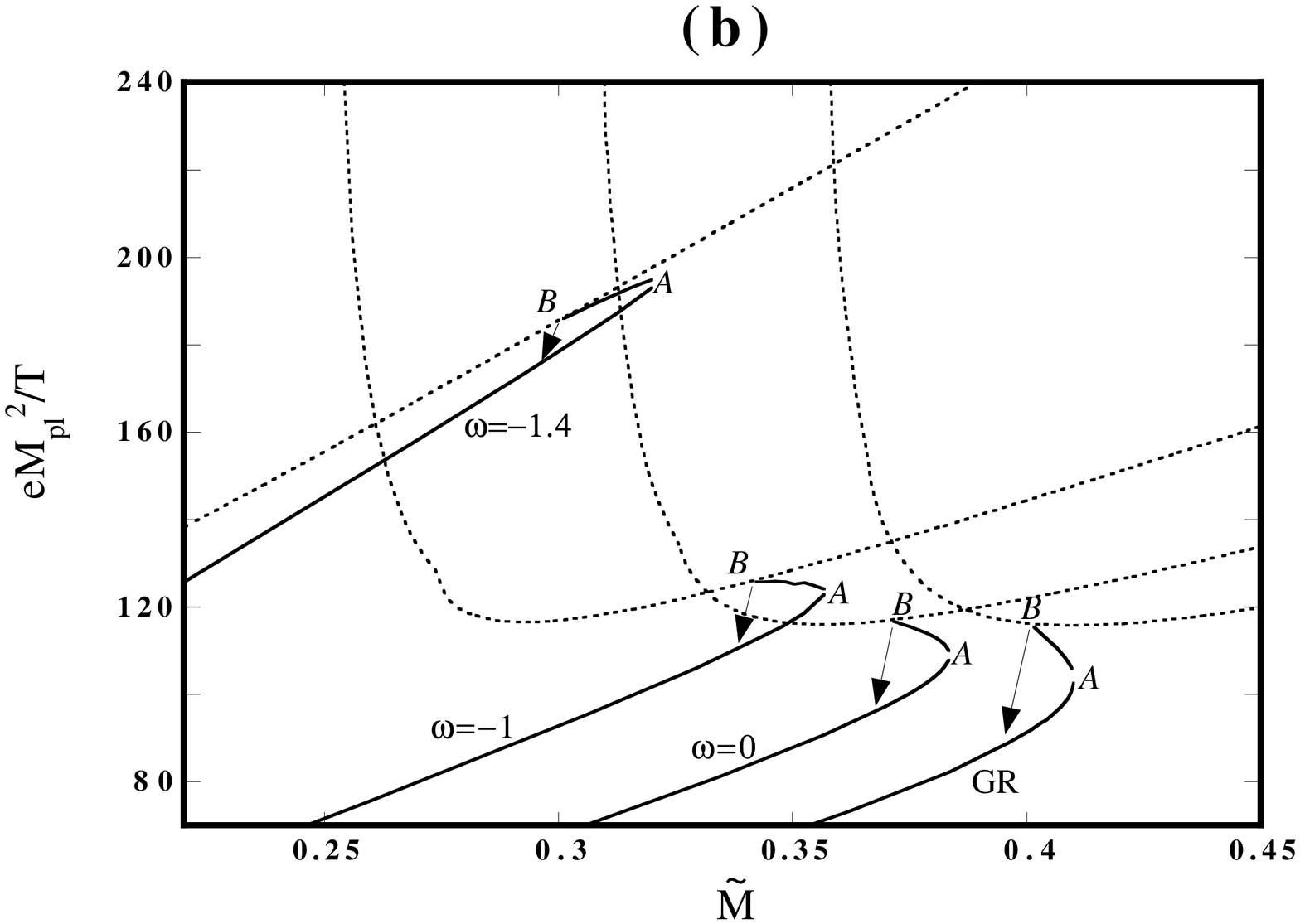}{}
\end{tabular}
\caption{  Mass-inverse temperature  diagram for (a) $\tilde{v}=0.1$, $\tilde{\lambda}=1$ and 
$\omega =0$, $\infty$, (b) $\tilde{v}=0.1$, $\tilde{\lambda}=0.1$ and 
$\omega =-1.4$, $-1$, $0$, $\infty$. 
When we consider the Hawking evaporation of a RN black hole,  
the solution experiences the second order phase transition for $\tilde{\lambda}=1$
when the solution changes to the monopole black hole at the point $B$.
On the contrary, it experiences first order phase transition  for $\tilde{\lambda}=0.1$.  
      \label{M-1Tv0.1comp}      }
\end{figure}
%%%%%%%%%%%%%%%%%%%%%%%%
%%%%%%%%%%%%%%%%%%%%
\begin{figure}
\begin{center}
\singlefig{12cm}{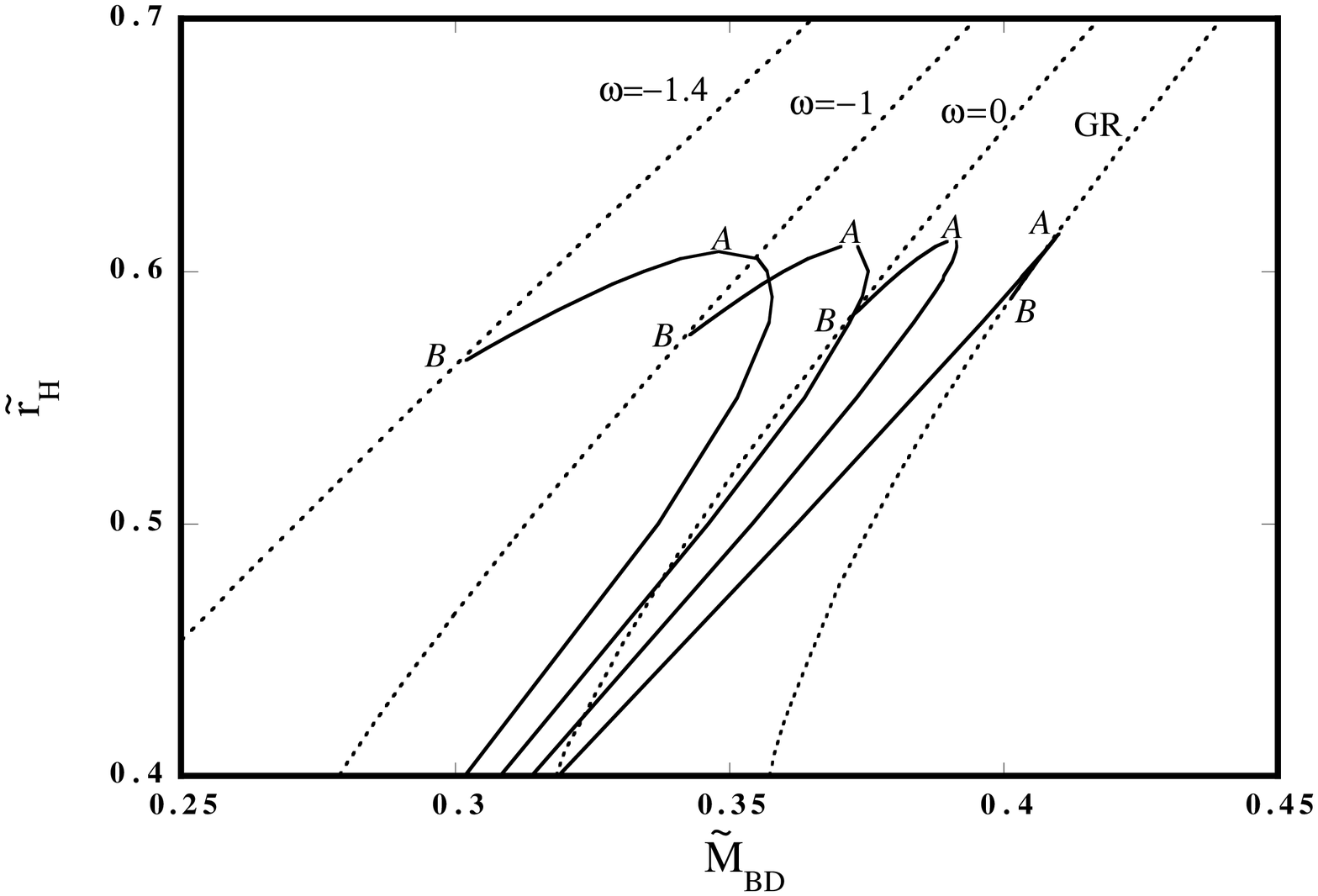}
\caption{Gravitational mass-horizon radius diagram of the monopole black hole in BD frame
for $\tilde{v}=0.1$, $\tilde{\lambda}=0.1$ and $\omega =-1.4$, $-1$, $0$, $\infty$. 
The solid lines denote the monopole black hole and the dotted lines denote the RN black hole. 
In this diagram, a cusp structure disappears. This may show that $\tilde{M}_{BD}$ is not 
appropriate to the control parameter of  catastrophe theory. 
\label{MBD-rhGR&BDv0.1l0.1}  }
\end{center}
\end{figure}
%%%%%%%%%%%%%%%%%%%%


\begin{thebibliography}{99}
\bibitem{BM}
R. Bartnik and J. Mckinnon, Phys. Rev. Lett. {\bf 61}, 141 (1988). 
\bibitem{Tor2} As a review paper see K. Maeda, Journal of the 
Korean Phys. Soc. {\bf 28}, S468, (1995), 
and M. S. Volkov and D. V. Gal'tsov, hep-th/9810070.
\bibitem{rf:colored}
M. S. Volkov and D. V. Galt'sov, Pis'ma Zh. Eksp. Theor. Fiz. {\bf 50}, 
312 (1989); P. Bizon, Phys. Rev. Lett. {\bf 64}, 2844 (1990); 
H. P. K\"{u}nzle 
and A. K. Masoud-ul-Alam, J. Math. Phys. {\bf 31}, 928 (1990).
\bibitem{Torii}
K. Maeda, T. Tachizawa, T. Torii and T. Maki, Phys. Rev.
Lett. {\bf 72}, 450 (1994);
T. Torii, K. Maeda and T. Tachizawa, Phys. Rev. D. {\bf 51},
1510 (1995).
\bibitem{DHS}
S. Droz, M. Heusler and N. Straumann, 
Phys. Lett. B {\bf 268}, 371 (1991);
P. Bizon and T. Chmaj, Phys. Lett. B {\bf 297}, 55 (1992).
\bibitem{LM}
H. Luckock and I. G. Moss, Phys. Lett. B {\bf 176}, 341 (1986); 
H. Luckock, in {\it String theory, quantum cosmology and quantum 
gravity, integrable and conformal invariant theories}, eds. H. de Vega 
and N. Sanchez, (World Scientific, Singapore, 1986), p. 455. 
\bibitem{Tor} T. Torii and K. Maeda, Phys. Rev. D {\bf 48}, 1643 
(1993).
\bibitem{rf:Greene}
B. R. Greene, S. D. Mathur and C. M. O'Neill, Phys. Rev. D {\bf 47}, 
2242 (1993).
\bibitem{Lee}
K. -Y. Lee, V. P. Nair and E. Weinberg, Phys. Rev. Lett. {\bf 68}, 1100 
(1992); Phys. Rev. D {\bf 45}, 2751 (1992); Gen. Relativ. Gravit. 
{\bf 24}, 1203 (1992)
\bibitem{Ortiz}
M. E. Ortiz, Phys. Rev. D. {\bf 45}, R2586 (1992). 
\bibitem{BFM}
P. Breitenlohner, P. Forg\'{a}cs and D. Maison, Nucl. Phys. B 
{\bf 383}, 357 (1992);  {\it ibid.} {\bf 442}, 126 (1995).
\bibitem{Aichelburg}
P. C. Aichelberg and P. Bizon, Phys. Rev. D. {\bf 48}, 607 (1993). 
\bibitem{Tachi}
T. Tachizawa, K. Maeda  and T. Torii, Phys. Rev. D.
{\bf 51},  4054 (1995). 
\bibitem{Donets}
E. E. Donets, D. V. Gal'tsov and M. Yu. Zotov, Phys. Rev. D {\bf 56}, 3459 (1997); 
{\it ibid.}, Pis'ma Zh. Eksp. Teor. Fiz., {\bf 65}, 855 (1997); P. Breitenlohner, G. 
Lavrelashvili and D. Maison, Nucl. Phys. B {\bf 524}, 427 (1998).
\bibitem{'t Hooft}
G. 't Hooft, Nucl. Phys. {\bf B 79}, 276 (1974); A. M. Polyakov, 
Pis'ma Zh. Eksp. Teor. Fiz. {\bf 20}, 430 (1974) [JETP Lett. {\bf 20}, 194 (1974)].
\bibitem{Sato}
A. H. Guth, Phys. Rev. D {\bf 23}, 347 (1981); J. R. Gott, Nature 
(London) {\bf 295}, 304 (1982); K. Sato, H. Kodama, M. Sasaki 
and K. Maeda, Phys. Lett. B {\bf 108}, 35 (1982). 
\bibitem{rf:Linde}
D. La and P.J. Steinhardt, Phys.
Rev. Lett. {\bf 62}, 376 (1989);
A.L. Berkin, K. Maeda and J. Yokoyama, {\it ibid.}
{\bf 65}, 141 (1990); A.L. Berkin and K. Maeda, Phys. Rev. D
{\bf 44}, 1691 (1991);
A. D. Linde, {\it ibid.} {\bf 49}, 748 (1994); J.
Garcia-Bellido, A. D. Linde and D. A. Linde, {\it ibid.} {\bf
50}, 730 (1994). 
\bibitem{Vilenkin}
A. Vilenkin, Phys. Rev. Lett. {\bf 72}, 3137 (1994);
A. D.  Linde, Phys. Lett. B {\bf 327}, 208 (1994);
N. Sakai, H. Shinkai, T. Tachizawa and K. Maeda, Phys. Rev. D {\bf 53}, 
655 (1996);
N. Sakai, {\it ibid.} {\bf 54}, 1548 (1996); I. Cho and A. Vilenkin, 
{\it ibid.} {\bf 56}, 7621 (1997). 
\bibitem{Sakai}
N. Sakai, J. Yokoyama and K. Maeda, Phys. Rev. D {\bf 59}, 103504 (1999).
\bibitem{GM}
G.W. Gibbons and K. Maeda, Nucl. Phys. B {\bf 298}, 741 (1988).
\bibitem{Hawking}
J.D.Bekenstein, Phys. Rev. D {\bf 5}, 1239 (1972);
S. W. Hawking, Comm. Math. Phys. {\bf 25}, 167 (1972).
\bibitem{Tamaki}
T. Tamaki, K. Maeda and T. Torii, Phys. Rev. D {\bf 57}, 4870 (1998). 
\bibitem{Hrad}
S. W. Hawking, Nature {\bf 248}, 30 (1974); Commun. Math. Phys. {\bf 43}, 
199 (1975). 
\bibitem{rf:MTW}
C. W. Misner, K. S. Thorne and J. A. Wheeler, {\it
Gravitation} (Freeman, New York 1973).
\bibitem{BD}
C. Brans and R. H. Dicke, Phys. Rev. {\bf 124}, 925 (1961).
\bibitem{Dicke}
R. H. Dicke, Phys. Rev. {\bf 125}, 2163 (1962). 
\bibitem{AD1}
Although singular solutions in the two conformal frames are not 
necessarily equivalent owing to the possibility of a 
singularity residing in the conformal rescaling (scalar field), 
solutions will map onto each other provided that they have a 
regular horizon and are asymptotically flat, which we assume here.
\bibitem{BPS}
M. K. Prasad and C. M. Sommerfield, Phys. Rev. Lett. {\bf 35}, 760 (1975). 
\bibitem{Maison}
D. Maison, Nucl. Phys. B. {\bf 182}, 144 (1981). 
\bibitem{Harvey}
J. A. Harvey and J. Liu, Phys. Lett. B {\bf 268}, 40 (1991).
\bibitem{CosmicString}
Similar phenomena can be seen in the cosmic string. There exist string solution and 
the Melvin solution under some value of $v$. As $v$ gets larger, both solution approaches 
each other. When it reaches critical value for which the deficit angle becomes $2\pi$, 
there exists only cylindrical solution\protect\cite{Christ}. 
\bibitem{Christ}
M. Christensen, A. L. Larsen and Y. Verbin, gr-qc/9904049.  
\bibitem{lambdabig}
For the very large value of $\tilde{\lambda}$, it is discussed in \protect\cite{Lue} that 
there appears a second minimum for $g^{rr}$ inside a first minimum near the $\tilde{v}_{extreme}$. 
As $\tilde{v}$ increase, the outer minimum moves upward, while the inner minimum downward. 
\bibitem{Lue}
A. Lue and E. J. Weinberg, hep-th/9905223.  
\bibitem{Barros}
A. Barros and C. Romero, Phys. Rev. D {\bf 56}, 6688 (1997);  
A. Banerjee, A. Beesham, S. Chatterjee and A. A. Sen, Class. Quantum Grav. {\bf 15},  
645 (1998); O. Dando and R. Gregory, Class. Quantum Grav. {\bf 15}, 985 (1998). 
\bibitem{Gundlach}
C. Gundlach and M. E. Ortiz, Phys. Rev. D {\bf 42}, 2521 (1990);  A. Barros and C. Romero, 
J. Math. Phys. {\bf 36}, 5800 (1995); A. A. Sen, N. Banerjee and A. Banerjee, Phys. Rev. D {\bf 56}, 
3706 (1997);M. Em\'{i}lia and X. Guimar\~{a}es, Class. Quantum Grav. {\bf 14}, 435 (1997). 
\bibitem{Damour}
T. Damour and A. Vilenkin, Phys. Rev. Lett. {\bf 78}, 2288 (1997); 
R. Gregory and C. Santos, Phys. Rev. D {\bf 56}, 1194 (1997). 
\bibitem{Dando}
O. Dando, gr-qc/9904058. 
\bibitem{note1}
The precise diagram is given in Ref.~\protect\cite{BFM}.
\bibitem{CAT}
T. Poston and I. Stewart, $Catastrophe\ Theory\ and\ Its   \
Applications,$ Pitman,  London (1978);
R. Thom, $Structure\ Stability\ and\ Morphogenesis,$ Benjamin
(1975).
\bibitem{Katz}
J. Katz, I. Okamoto and O. Kaburaki, Class. Quantum Grav. {\bf
10}, 1323  (1993).
\bibitem{Will}
C. Will, $Theory\ and\ experiment\ in\ gravitational\ physics$, 
(Cambridge university press, Cambridge 1981).
\bibitem{footnote1}
It was shown that gravitational mass 
in the Einstein frame satisfy the first law of black hole thermodynamics 
\protect\cite{Koga}.
\bibitem{Koga}
J. Koga and K. Maeda, Phys. Rev. D {\bf 58} 064020 (1998). 
\bibitem{Callan}
C. G. Callan and J. M. Maldacena, Nucl. Phys. B {\bf 513}, 198 (1998); G. Gibbons, 
Nucl. Phys. B {\bf 514}, 603 (1998). 
\bibitem{Grandi}
N. Grandi, E. F. Moreno and F. A. Schaposnik, hep-th/9901073; P. K. Tripathy, hep-th/9904186. 
\end{thebibliography}
\end{document}